\documentclass[11pt,a4paper]{article}
\usepackage[T1]{fontenc}
\usepackage{lmodern}
\usepackage[margin=25mm]{geometry}
\usepackage{amsmath,amssymb,amsthm,mathtools}
\usepackage{booktabs,array,tabularx,graphicx,tikz,microtype}
\usepackage{float}
\usepackage[numbers,square,sort&compress]{natbib}
\usepackage[hidelinks]{hyperref}
\usepackage{cleveref}

\newtheorem{theorem}{Theorem}
\newtheorem{lemma}{Lemma}
\newtheorem{proposition}{Proposition}

\theoremstyle{definition}
\newtheorem{definition}{Definition}
\theoremstyle{remark}

\floatstyle{ruled}
\newfloat{algorithm}{tbp}{loa}
\floatname{algorithm}{Algorithm}
\newcounter{pseudocode}
\newcommand{\pseudoline}[2]{\stepcounter{pseudocode}\thepseudocode&\hspace*{#1em}#2\\}
\crefname{equation}{}{}
\Crefname{equation}{}{}
\creflabelformat{equation}{#2(#1)#3}
\newcommand{\Appendixref}[1]{Appendix~\ref{#1}}

\newcommand{\N}{\mathcal N}
\newcommand{\G}{\mathcal G}
\newcommand{\eps}{\varepsilon}
\newcommand{\OO}{\mathcal O}
\newcommand{\OPT}{\mathrm{OPT}}
\newcommand{\F}{\mathcal F}

\title{A $(3/2+\varepsilon)$-Approximation for Weighted 3-Set Packing}
    \author{Changjun Wang\textsuperscript{a}\thanks{Email: \texttt{wcj@amss.ac.cn}.}\qquad Daorong Yu\textsuperscript{a}\thanks{Email: \texttt{yudaorong@amss.ac.cn}.} \qquad Wei Yu\textsuperscript{b}\thanks{Email: \texttt{yuwei@ecust.edu.cn}.}\\[5pt]
\small  \textsuperscript{a}SKLMS, Academy of Mathematics and Systems Science, Chinese Academy of Sciences, Beijing, China \\
\small  \textsuperscript{b}School of Mathematics, East China University of Science and Technology, Shanghai, China}
\date{}
\hypersetup{pdfauthor={Wang, Yu, Yu},
 pdftitle={A (3/2+epsilon)-Approximation for Weighted 3-Set Packing},
 pdfsubject={Approximation by power local search and gain recovery}}
\begin{document}
\maketitle
\begin{abstract}
We give a deterministic $(3/2+\varepsilon)$-approximation for weighted
$3$-set packing, for every fixed rational $\varepsilon>0$,  improving
on the 1.6908-approximation reported by Tong and Xu (2026). The algorithm has two phases. It first performs bounded local search using a rounded $12/7$-power
potential. It then combines compatible exchanges that insert at most
three sets and increase the original weight. The analysis relates the
gap left by the first phase to the gain available to the second phase.
We prove a structural inequality that bounds the weight of a comparison
packing by $3/2$ times the current weight plus one sixth of a sum of
small-exchange gains. Each gain accounts for all sets that the exchange
must delete. A greedy rule recovers enough of these gains to obtain
the approximation guarantee. We derive the structural bound from local
optimality by analyzing finite computation trees and averaging the
error caused by their limited depth. The running time is
$n^{4^{O(1/\varepsilon)}}\operatorname{poly}(L)$, where $n$ is the
number of input sets and $L$ is the input encoding length. The same
guarantee holds for maximum-weight independent set in induced
$K_{1,4}$-free graphs.
\end{abstract}

\section{Introduction}\label{sec:introduction}
Weighted $k$-set packing asks us to choose a collection of disjoint
sets of maximum total weight. An instance consists of a finite ground
set $Z$, an explicitly listed family $\F\subseteq 2^Z$ with
$|S|\le k$ for every $S\in\F$, and weights
$w:\F\to\mathbb Q_{\ge0}$. A \emph{packing} is a subfamily
$A\subseteq\F$ whose sets are pairwise disjoint. We write
$w(Y)=\sum_{S\in Y}w(S)$ for any $Y\subseteq\F$, and seek a packing
$A$ maximizing $w(A)$. In the resource-allocation interpretation,
each element is a resource that can be used once, and each set is
a choice that uses those resources and earns its weight. Equivalently,
the problem is maximum-weight matching in a hypergraph of rank at
most $k$. A direct application of the case $k=3$ is cycle-based kidney exchange. Each feasible exchange cycle involves a group of patient--donor pairs and has an associated benefit. When only cycles of length two or
three are allowed, maximizing the total benefit of disjoint exchanges
is a weighted $3$-set packing problem~\cite{BMR09}.

The case $k=3$ is the first difficult case beyond graph matching.
Weighted $2$-set packing can be solved in polynomial time, whereas
$3$-set packing contains $3$-dimensional matching and is NP-hard even
with unit weights. The unit-weight problem also admits no polynomial-time
approximation scheme unless $P=NP$~\cite{CC06}.
At the same time, sets of size at most three impose a useful restriction
on conflicts. Any one set can intersect at most three members of
a packing. This restriction supports approximation algorithms that
improve a packing through small exchanges.

Local search starts with a packing, inserts a small family of disjoint
sets, and deletes the current sets that intersect them. For unit weights,
an exchange improves the objective whenever it inserts more sets than
it deletes. Bounded-exchange analysis goes back to Hurkens and
Schrijver~\cite{HS89}, and a $(4/3+\varepsilon)$-approximation is known
for unweighted $3$-set packing~\cite{Cyg13,fureryu2014}.
Unequal weights make this comparison harder. Three new sets can be
less valuable than one current set, and a useful exchange may need
to combine gains and losses at several places.

Early algorithms for the weighted problem used exchanges that increase
total weight~\cite{AH98,CH01}. Berman~\cite{berman2000} introduced
local search based on the sum of squared weights, obtaining a
$((k+1)/2+\varepsilon)$-approximation and hence a
$(2+\varepsilon)$-approximation for $k=3$. Giving heavier sets more
influence in the search made it possible to prove a stronger weight
bound. Neuwohner~\cite{Neu21} later improved on this guarantee.
For $k=3$, Thiery and Ward~\cite{thieryward2023} achieved a ratio of
$1.786$ by analyzing larger exchanges of constant size.
Thiery's layered analysis~\cite{thiery2023thesis} subsequently approached
$\sqrt3$ and gave a matching limitation for squared-weight local search.
Tong and Xu~\cite{tongxu2026} reported a $1.6908$-approximation by adding
a second phase that recovers gains in the original weight.
For larger $k$, unweighted subroutines~\cite{Neu23} and ordered local
search~\cite{ST26} give further improvements. On the hardness side, Hazan, Safra, and Schwartz~\cite{HSS06} proved an $\Omega(k/\log k)$ inapproximability bound unless $P=NP$. Lee, Svensson, and Thiery~\cite{LST25} obtained a linear-in-$k$ bound unless $\mathsf{NP}\subseteq\mathsf{BPP}$. These bounds concern growing $k$; our focus is the fixed case $k=3$.

Using a power potential leaves an opportunity for such a second phase.
For example, one set of weight $16$ can intersect three mutually disjoint
sets of weight $6$ each. Replacing the former by the latter increases
total weight from $16$ to $18$, while decreasing the $12/7$-power
potential. Thus a packing that is locally optimal for the potential
may still admit useful exchanges in the original weight.
To exploit this opportunity, we need a bound that relates the gap to
an optimal packing to these remaining gains. We must also account for
conflicts between proposed exchanges and for the full cost of the sets
that each exchange deletes.

\subsection{Our result}
We obtain the following approximation guarantee. As usual, a
$\rho$-approximation for this maximization problem returns a packing
of weight at least $1/\rho$ times the optimum. Here $n$ denotes the
number of input sets, or graph vertices in the graph extension, and
$L$ is the binary input encoding length.

\begin{theorem}\label{thm:algorithm}
For every fixed rational $0<\varepsilon\le1$, there is a deterministic
$(3/2+\varepsilon)$-approximation for weighted $3$-set packing in time
$n^{4^{O(1/\varepsilon)}}\operatorname{poly}(L)$. The first phase inserts
at most $4^{\lceil1/\varepsilon\rceil}-1$ sets per exchange, and the
second phase enumerates $O(n^3)$ insertions of at most three sets.
The same guarantee and running-time bound hold for maximum-weight
independent set in induced $K_{1,4}$-free graphs.
\end{theorem}

The first phase uses a rounded $12/7$-power potential.
The second phase freezes the resulting packing and considers all
positive-gain insertions of at most three sets in order of decreasing
gain. It accepts an insertion when its sets are disjoint from all
previously accepted insertions. The algorithm is polynomial for
each fixed $\varepsilon$, although its first-phase neighborhood grows
rapidly as $\varepsilon$ decreases.

The two-phase design and the view of gain recovery as a packing
problem build on Tong and Xu~\cite{tongxu2026}. Our main contribution
is a structural inequality that keeps track of both the current
weight and the gain available to recovery. A construction using
finite computation trees derives this inequality from bounded local
optimality. Its coefficients allow a greedy recovery rule to prove
the improved ratio.

\subsection{How the two phases fit together}
Consider the preprocessed instance with positive weights and temporarily
ignore rounding. Let $A$ be the first-phase packing and $O$ an optimal
packing, with common sets removed from both. Join $a\in A$ and $o\in O$
when they intersect. This bipartite comparison graph has maximum degree
three. Every $o\in O$ has a neighbor in $A$, since otherwise adding
$o$ alone would improve the potential.

Assign each $o\in O$ to one of its neighbors in $A$. For $a\in A$,
let $C_a$ contain the sets assigned to $a$. The groups partition $O$,
and each contains at most three disjoint sets. Inserting $C_a$ requires
deleting every set in $A$ that intersects any member of $C_a$.
Write $\N_A(C_a)$ for this full deletion neighborhood. The gain from
inserting the group on its own is therefore
$w(C_a)-w(\N_A(C_a))$. For a fixed assignment, sum the positive
group gains. Let $\G_w(A,O)$ be the largest such sum over all assignments.
These assignments allow us to compare the two packings in the proof.
The second phase enumerates all small insertions, so its candidate
list includes every positive-gain group.

For an integer $H\ge1$, we prove that local optimality under insertions
of at most $4^H-1$ sets implies
\[
 w(O)\le\left(\frac32+\frac7{8H}\right)w(A)
          +\frac16\G_w(A,O).
\]
The additional term measures how much gain the comparison packing
makes available to the recovery phase. Let $\Delta$ be the sum of
the gains declared by the insertions accepted in that phase, and
let $B$ be its output. Greedy selection gives
\[
 \G_w(A,O)\le9\Delta,
 \qquad w(B)\ge w(A)+\Delta.
\]
The factor nine follows from the number of comparison groups that one
accepted insertion can meet. Shared deletions can only increase the
final improvement; Section~\ref{sec:architecture} gives the full argument.
Consequently,
\[
 w(O)\le\left(\frac32+\frac7{8H}\right)w(A)+\frac32\Delta
       \le\left(\frac32+\frac7{8H}\right)w(B).
\]
The product $(1/6)\cdot9=3/2$ explains the gain coefficient in our
structural bound. Choosing $H=\lceil1/\varepsilon\rceil$ leaves
enough room for the small rounding loss.

The structural argument has two parts. First, we split each current
set's power weight into amounts transferred to neighboring comparison
sets and a residual amount. Under suitable constraints on these amounts,
we prove the desired weight bound by charging each comparison set
to current sets. A group whose charge is reduced by one sixth must
also reserve one sixth of its full deletion cost. The resulting
reduction is exactly one sixth of that group's gain.
Local estimates prove that these charges fit within $3/2$ times
each current weight.

Second, we construct the required transfers from local optimality.
We unfold the comparison graph into trees so that gains can be computed
recursively along branches. A profitable tree exchange yields a
profitable exchange in the original graph, even if a set has several
copies in the tree. This observation bounds the branch gains.
Stopping the recursion at a finite depth leaves a small deficit in
the constraints. We cover it by adding part of the next level's gains.
These additions sum to at most the original power weight when summed
over depths. Averaging over $H$ depths gives the term $7/(8H)$.

The exponent $12/7$ is a choice for which the residual credits,
H\"older bounds, and threshold estimates in this analysis all hold
together. Other exponents may also permit the same guarantee; the
examples in Appendix~\ref{sec:exponent-examples} exclude only some ranges.
A separate example in Appendix~\ref{sec:framework-lower-bound} shows
that a guarantee covering every permitted first-phase choice in this
framework cannot improve on $3/2$. Section~\ref{sec:scope} discusses
the scope of these limitations.

Section~\ref{sec:algorithm} gives the model and algorithm.
Section~\ref{sec:architecture} proves the gain-recovery bound and
sets out the remaining proof. Sections~\ref{sec:mixed-proof}
and~\ref{sec:finite} establish the structural bound and derive its
hypotheses from local optimality.
Section~\ref{sec:rounding} completes the approximation and running-time
analysis, and Section~\ref{sec:scope} gives the graph extension and
discusses limitations. The appendices contain the remaining scalar
cases, derivative calculations, and lower-bound constructions.

\section{Model and algorithm}\label{sec:problem}\label{sec:algorithm}
We work with the weighted $3$-set packing instance $(Z,\F,w)$ defined
in the introduction. A factor $\rho\ge1$ means that the returned packing
$B$ satisfies $w(O^\star)\le\rho w(B)$ for an optimal packing $O^\star$.
We first describe the notation for an exchange, then give the two phases
of the algorithm.

\subsection{Preprocessing and exchanges}
Discard zero-weight sets and put every positive-weight empty set in a
packing $P$. These empty sets can be included in every solution.
If no sets remain, return $P$. Otherwise, reuse $\F,w$ for the remaining
instance, whose sets are nonempty and positively weighted, and put
$n=|\F|$. Let $L$ be the binary encoding length of the original input.
We analyze the reduced instance and add $P$ to the final output.
This preserves any approximation factor at least one.

For auxiliary weights $v:\F\to\mathbb R_{\ge0}$, use the additive notation
$v(Y)=\sum_{S\in Y}v(S)$. The analysis will use weights $v$ obtained
from rounding the input weights $w$.
For a current packing $A$ and a packing $X\subseteq\F\setminus A$, define
\[
 \N_A(X)=\{a\in A:a\cap S\ne\varnothing\text{ for some }S\in X\},
 \qquad g_A^v(X)=v(X)-v(\N_A(X)).
\]
The exchange inserts $X$ and removes $\N_A(X)$, producing the packing
$(A\setminus\N_A(X))\cup X$. Thus $g_A^v(X)$ is its change in total
$v$-weight. Write $g_A=g_A^w$ for the change in the original objective.
Each set in $X$ has at most three elements, and each element belongs
to at most one member of $A$. Hence $|\N_A(X)|\le3|X|$.

For $p>0$, let $\Phi_p^v(Y)=\sum_{S\in Y}v(S)^p$.
A packing $A$ is \emph{$b$-locally optimal} for this potential if
$\Phi_p^v(X)\le\Phi_p^v(\N_A(X))$ for every packing
$X\subseteq\F\setminus A$ with $|X|\le b$.
This condition says that none of the allowed exchanges improves the
potential. We use
\[
 p=\frac{12}{7},\qquad r=\frac1p=\frac7{12}
\]
throughout the algorithm and its analysis.

For packings $A,O$ that are disjoint as families, the \emph{comparison
graph} has parts $A,O$ and an edge $ao$ exactly when $a\cap o\ne\varnothing$.
We write $a\sim o$ for adjacency and $\N(u)$ for the set of neighbors
of any vertex $u$ in this graph. Both sides have degree at most three,
by the same element-counting argument. The notation $\N_A(X)$ also
denotes the neighborhood of $X\subseteq O$ in this graph.
If two packings share a subfamily $C=A\cap O$, then no set in
$O\setminus C$ meets a member of $C$. Removing common members therefore
leaves the deletion neighborhood of every $X\subseteq O\setminus C$
unchanged. We use this observation when comparing the algorithm with
an optimum.

\subsection{Power local search followed by gain recovery}
The first phase repeatedly improves a rounded power potential.
Rounding makes the potential integer-valued and bounds the number of
iterations. The second phase uses the original weights to select small
improvements relative to the final first-phase packing.

For a rational accuracy parameter $0<\eps\le1$, set
\begin{equation}
 \begin{gathered}
 H=\lceil1/\eps\rceil,\qquad b=4^H-1,\qquad M=\max_{S\in\F}w(S),\\
 \tau=\frac{\eps M}{30n},\qquad
 Q(S)=\left\lfloor(w(S)/\tau)^p\right\rfloor.
 \end{gathered}
 \label{eq:parameters}
\end{equation}
Here $H$ is the depth used in the analysis, $b$ is the first-phase
insertion bound, and $\tau$ is the rounding scale. Extend $Q$ additively
to subfamilies. Section~\ref{sec:rounding} explains why this scale leaves enough room for the rounding error and how to compute $Q$ exactly. For $\eps>1$, it suffices to run the algorithm with accuracy one.

To connect the integer potential with the analysis, define the auxiliary weights
\[
 v(S)=\tau Q(S)^r\qquad(S\in\F).
\]
Since $v(S)^p=\tau^pQ(S)$, an exchange increases $Q$ if and only if it increases $\Phi_p^v$. Thus the first phase is also a local search for the power potential $\Phi_p^v$. The weights $v$ are used only in the analysis; the algorithm compares the integer values $Q$.

\begin{algorithm}[htbp]
\caption{Power local search and gain recovery}\label{alg:two-stage}
\setcounter{pseudocode}{0}
\begin{tabularx}{\linewidth}{@{}r@{\quad}X@{}}
\pseudoline{0}{$A\gets\varnothing$.}
\pseudoline{0}{\textbf{while} some packing $X\subseteq\F\setminus A$, $1\le |X|\le b$, has $Q(X)>Q(\N_A(X))$ \textbf{do}}
\pseudoline{1}{$A\gets(A\setminus\N_A(X))\cup X$.}
\pseudoline{0}{\textbf{end while}; freeze $A$.}
\pseudoline{0}{List all packings $X\subseteq\F\setminus A$ with $1\le |X|\le3$ and $g_A(X)>0$.}
\pseudoline{0}{Order this list by nonincreasing $g_A(X)$; set $U\gets\varnothing$, $\mathcal R\gets\varnothing$.}
\pseudoline{0}{\textbf{for} each candidate $X$ in this order \textbf{do}}
\pseudoline{1}{\textbf{if} $S\cap T=\varnothing$ for all $S\in X$, $T\in U$, \textbf{then} $U\gets U\cup X$ and $\mathcal R\gets\mathcal R\cup\{X\}$.}
\pseudoline{0}{\textbf{end for}.}
\pseudoline{0}{\textbf{return} $B=(A\setminus\N_A(U))\cup U$, together with $P$.}
\end{tabularx}
\end{algorithm}

Fixed total orders resolve all choices and ties, making the procedure
deterministic. Sets with $Q(S)=0$ may be omitted from first-phase
insertions. They remain available for recovery.
In the second phase, every candidate's gain is measured against the
same frozen packing $A$. The chosen insertions must be mutually disjoint,
but their deletion neighborhoods may overlap. Each removed set is
deleted only once when the output is formed.
Let
\begin{equation}
 \Delta=\sum_{X\in\mathcal R}g_A(X)
 \label{eq:accepted-gain}
\end{equation}
be the sum of the individually computed gains.
Overlapping deletions can make the actual improvement larger than
$\Delta$. Sorting by gain also ensures that every rejected candidate
meets an accepted candidate of at least as large a gain. These two
facts give the recovery bound in the next section.

\section{Gain recovery and the proof plan}\label{sec:architecture}
We now describe the gain that the structural analysis will provide
and prove that the second phase recovers a fixed fraction of it.
This step uses only the size of the sets and the ordering of candidates.

\subsection{Grouping the comparison packing}
Let $A,O$ be packings disjoint as families, and suppose every $o\in O$
has a neighbor in $A$. For positive comparison weights this follows
from local optimality under single-set insertions.
Assign each $o$ to one adjacent member of $A$.
A group then consists of the sets assigned to the same current set.
Each group is a packing of at most three sets, so it has the size of
a candidate considered by recovery.

\begin{definition}\label{def:assignment}
An assignment $\pi$ maps each $o\in O$ to a neighbor in $A$ and forms
groups $C_a=\{o\in O:\pi(o)=a\}$. For nonnegative weights $v$, define
the \emph{assignment gain} and its maximum over all assignments by
\begin{equation}
 G_\pi^v=\sum_{a\in A}[v(C_a)-v(\N_A(C_a))]_+,
 \qquad \G_v(A,O)=\max_\pi G_\pi^v,
 \label{eq:assignment-gain}
\end{equation}
where $[x]_+=\max\{x,0\}$.
\end{definition}

Inserting a nonempty $C_a$ requires removing every set in $A$ that
meets it. This includes $a$ and may include other current sets.
The subtraction of $v(\N_A(C_a))$ accounts for all of these deletions.
The positive part retains the groups that improve the objective.
Assignments organize the proof. The algorithm obtains its candidates
by enumerating small packings in the input.

\paragraph{An example with a shared deletion.}
Let $A=\{a_0,a_1,a_2\}$ have weights $1,4,4$. For $i=1,2$, let $C_i$
contain three sets of weight $2$. All three meet $a_i$, and exactly
one also meets $a_0$. The six comparison sets are pairwise disjoint,
and there are no other intersections. These incidences can be realized
by sets of size at most three, by using one ground element per edge
of the comparison graph.
Every nonempty insertion from $C_i$ removes $a_i$, and
$3\cdot2^{12/7}<4^{12/7}$. Thus $A$ is locally optimal for the
unrounded power potential on this instance.
Each full group has gain $6-(4+1)=1$.
Both groups can be selected. Their declared gains sum to $2$, whereas
the resulting improvement is $12-(4+4+1)=3$, because $a_0$ is removed once.

\subsection{The greedy recovery bound}
\begin{lemma}\label{lem:greedy}
For every comparison packing $O$ as in Definition~\ref{def:assignment},
Algorithm~\ref{alg:two-stage} satisfies
\[
 \G_w(A,O)\le9\Delta,\qquad w(B)\ge w(A)+\Delta.
\]
\end{lemma}
\begin{proof}
Fix an assignment. Every positive-gain group is a recovery candidate.
Assign its gain to the first accepted candidate that intersects it,
including itself if it is accepted. Such a candidate exists because
otherwise the group would be accepted when considered.
The ordering of candidates ensures that the accepted candidate's gain
is at least the group's gain.

The comparison groups are subfamilies of the packing $O$ and use
pairwise disjoint ground elements. An accepted candidate contains at
most three sets of size at most three. Its at most nine elements
therefore meet at most nine groups. It follows that
$G_\pi^w\le9\Delta$. Taking the maximum over assignments gives the first
inequality.

For the second inequality, the accepted insertions form a packing $U$,
and $\N_A(U)$ is the union of their deletion neighborhoods. Since
weights are nonnegative,
\[
 \begin{aligned}
 w(B)-w(A)
 &=w(U)-w(\N_A(U))\\
 &\ge\sum_{X\in\mathcal R}
       \bigl(w(X)-w(\N_A(X))\bigr)=\Delta.
 \end{aligned}
\]
A shared deletion contributes several times to the declared costs
and only once to the actual cost.
\end{proof}

\subsection{The estimate needed from local optimality}
For the moment, ignore rounding and common members. Put
$\alpha_H=3/2+7/(8H)$. Sections~\ref{sec:mixed-proof} and~\ref{sec:finite}
will prove
\[
 w(O)\le\alpha_Hw(A)+\frac16\G_w(A,O)
\]
whenever $A$ is locally optimal for the $12/7$-power potential under
insertions of size at most $4^H-1$.
Lemma~\ref{lem:greedy} then gives
\[
 w(O)\le\alpha_Hw(A)+\frac32\Delta
       \le\alpha_H\bigl(w(A)+\Delta\bigr)
       \le\alpha_Hw(B).
\]
Thus the role of the next two sections is to bound the part of the
comparison weight left beyond $\alpha_Hw(A)$ by the gains of small
groups. Section~\ref{sec:mixed-proof} establishes such a bound from
a decomposition of the power weights. Section~\ref{sec:finite}
constructs the decomposition from local optimality, with the error
$7/(8H)$. Section~\ref{sec:rounding} includes rounding and common
members to finish the approximation proof.

\section{A mixed charging inequality}\label{sec:mixed-proof}
The estimate in this section connects the two phases of the algorithm.
It bounds the weight of a comparison packing by the current weight and
the gain available from groups of at most three sets. We first assume
that the power weights admit a suitable decomposition. Section~\ref{sec:finite}
constructs this decomposition from local optimality, with a small increase
in the current budgets.

Let $A,O$ be the two parts of a finite simple bipartite graph of maximum
degree three, with no isolated vertex in $O$. For $X\subseteq O$, let
$\N(X)\subseteq A$ be its full neighborhood. Fix $r=7/12$.
Each current vertex $a\in A$ has a budget $c_a$, divided into transfers $F_{oa}$ to its neighbors and a remaining amount $t_a$, called its
\emph{residual}. Each comparison vertex $o\in O$ has a demand $q_o$.

In the application in Section~\ref{sec:finite}, the demands are $q_o=v(o)^p$, where $p=12/7$. The current budgets are obtained from $v(a)^p$ by adding nonnegative amounts. These additions make the constraints below hold, and their effect on the final weight bound is controlled by averaging over depths.

We first work with arbitrary nonnegative budgets, demands, transfers,
and residuals satisfying
\begin{align}
 c_a&=t_a+\sum_{o\sim a}F_{oa},\label{eq:budget}\\
 q_o&\le\sum_{a\sim o}F_{oa}+\sum_{a\sim o}t_a-\max_{a\sim o}t_a.
 \label{eq:demand}
\end{align}
Thus a demand can use all incoming transfers and all neighboring
residuals except the largest one. Equivalently, the residual contribution
is $\min_{a\sim o}\sum_{b\sim o,\,b\ne a}t_b$. This form allows any
one neighbor to serve as a parent whose residual is excluded, as required
by the branch construction in Section~\ref{sec:finite}.

Set $w_a=c_a^r$ and $w_o=q_o^r$, and extend $w$ additively to vertex sets.
These are the weights associated with the budgets and demands.
An assignment chooses a neighbor $\pi(o)$ for each $o\in O$ and forms
$C_a=\{o:\pi(o)=a\}$. Following Definition~\ref{def:assignment}, write
\[
 \G=\max_\pi\sum_{a\in A}[w(C_a)-w(\N(C_a))]_+.
\]

\begin{theorem}\label{thm:mixed}
Under \eqref{eq:budget}--\eqref{eq:demand},
\begin{equation}
 w(O)\le\tfrac32w(A)+\tfrac16\G.\label{eq:theorem}
\end{equation}
\end{theorem}

Two simple cases explain the terms in this bound. If all transfers
vanish, then $c_a=t_a$ and
$q_o^r\le\frac12\sum_{a\sim o}t_a^r$. For three neighboring residuals
$t_1\ge t_2\ge t_3$, this follows from
\[
 (t_2+t_3)^r\le t_2^r+(2^r-1)t_3^r
 \le\tfrac12(t_1^r+t_2^r+t_3^r).
\]
The first inequality follows because $(x+1)^r-x^r$ decreases for $x\ge1$.
The second uses $t_1\ge t_2$ and $2^r\le3/2$. With two neighbors,
the demand is at most the smaller residual, and with one neighbor it
is zero. Summing over $O$ gives $3w(A)/2$, since each current vertex
has at most three neighbors.

Transfers can support more comparison weight. A star with one current
vertex of budget $1$, residual $0$, and three transfers of size $1/3$
supports three demands of size $1/3$. Their total weight is
$3^{5/12}>3/2$. The three leaves also form a group with gain
$3^{5/12}-1$. The second term of \eqref{eq:theorem} keeps track of
such gains for the recovery phase.

\subsection{The bound to be proved}
We will choose an assignment and a set $S\subseteq A$ of owners whose
groups receive a discount. For these groups, we count $5/6$ of the
comparison weight and add $1/6$ of the full deletion cost. The required
bound is
\begin{equation}
 \sum_{a\notin S}w(C_a)+\tfrac56\sum_{a\in S}w(C_a)
 +\tfrac16\sum_{a\in S}w(\N(C_a))\le\tfrac32w(A).
 \label{eq:certificate}
\end{equation}
Its left-hand side is exactly
\[
 w(O)-\tfrac16\sum_{a\in S}\bigl(w(C_a)-w(\N(C_a))\bigr).
\]
The sum being subtracted is at most $\G$, because taking positive parts
and then maximizing over assignments can only increase it.
Consequently, \eqref{eq:certificate} implies Theorem~\ref{thm:mixed}.
To prove it, we distribute its charges among current vertices and bound
the charge at each $a$ by $3w_a/2$.

\subsection{Choosing owners and splitting comparison weights}
Keep the budgets $c_a$ and demands $q_o$ fixed, and minimize the total
transfer $\sum_{oa}F_{oa}$ over all decompositions satisfying
\eqref{eq:budget}--\eqref{eq:demand}. A minimum exists because the
feasible set is nonempty and compact. At a minimum, every positive
transfer $F_{oa}$ comes from a largest residual adjacent to $o$, and
$o$'s demand constraint is tight.

To see this, decrease $F_{oa}$ by a small $\delta>0$ and increase $t_a$
by $\delta$. The budget at $a$ is unchanged. At every other comparison
neighbor, the residual sum increases by $\delta$ while its maximum
increases by at most $\delta$, so the demand constraint remains valid.
At $o$, the sum of transfers and residuals is unchanged. If $t_a$ is
smaller than the largest residual, choose $\delta$ small enough that
the maximum remains unchanged. If the demand constraint has slack,
choose $\delta$ smaller than that slack. Either case would allow a
smaller total transfer, contradicting minimality.

Choose $\pi(o)$ along a largest positive transfer, or along a largest
residual if all transfers to $o$ vanish. Call $\pi(o)$ the \emph{owner}
of $o$. The owner's residual is therefore always largest among the
neighbors of $o$. Define
\begin{equation}
 P_o=\left(F_{o,\pi(o)}+\sum_{b\sim o,\,b\ne\pi(o)}t_b\right)^r,
 \qquad J_t(x)=(t+2x)^r-(t+x)^r.
 \label{eq:maincharge}
\end{equation}
The term $P_o$ uses the owner's transfer and the available residuals.
The function $J_t$ bounds the extra weight supported by a transfer
from another neighbor. More precisely,
\begin{equation}
 w_o\le P_o+\sum_{b\sim o,\,b\ne\pi(o)}J_{t_b}(F_{ob}).
 \label{eq:split}
\end{equation}
Indeed, for an off-owner transfer $x=F_{ob}>0$, the quantity inside
$P_o$ already contains $t_b$ and an owner transfer at least $x$.
Since $u^r$ is concave, adding $x$ increases its value by at most
$(t_b+2x)^r-(t_b+x)^r$. Further transfers have still smaller marginal
increases. Summing these bounds proves \eqref{eq:split}. If all
transfers vanish, the same inequality follows from \eqref{eq:demand}.

\subsection{Local charges and deletion reserves}
Add zero-demand, zero-transfer leaves until every current vertex has
degree three. Each new leaf is assigned to its only neighbor. This
leaves $\G$ unchanged. An originally nonempty group already has its
owner in its deletion neighborhood, and an originally empty group
still has nonpositive gain after receiving zero-weight leaves.
We make this addition before choosing the discounts.

Let $k_a$ be the number of comparison vertices owned by $a$. Put $a\in S$
exactly when
\[
 \bigl[k_a=2,\ c_a\ge(7/3)^{12/5}t_a\bigr]
 \quad\text{or}\quad
 \bigl[k_a=3,\ c_a\ge(8/3)^{12/5}t_a\bigr],
\]
with $c_a=t_a=0$ left ordinary. We call an owner in $S$ \emph{discounted}
and any other owner \emph{ordinary}. Set
\[
 (s_a,\lambda_a)=
 \begin{cases}
 (1,1/2),&a\notin S,\\
 (5/6,7/24),&a\in S.
 \end{cases}
\]
Here $s_a$ is the fraction of the comparison charge counted for $C_a$.
The coefficient $\lambda_a$ specifies a residual credit, which
redistributes charges between current vertices and cancels in their sum.
The calculation below explains how this credit determines the thresholds.

For each comparison vertex $o$ owned by $a$, charge $s_aP_o$ to $a$.
For every other neighbor $b$ of $o$, give $a$ a credit
$\lambda_a t_b^r$ and charge the same amount to $b$. Also charge $b$
for its increment $s_aJ_{t_b}(F_{ob})$. Finally, reserve one sixth of
the weight of each current vertex that may have to be deleted for a
discounted group. The resulting account at $a$ is
\begin{align}
 E_a={}&s_a\sum_{o\in C_a}P_o
 -\lambda_a\sum_{o\in C_a}\sum_{b\sim o,\,b\ne a}t_b^r\notag\\
 &+\sum_{o\sim a,\,\pi(o)\ne a}
    \bigl(s_{\pi(o)}J_{t_a}(F_{oa})+\lambda_{\pi(o)}t_a^r\bigr)\notag\\
 &+\frac{w_a}{6}\left(\boldsymbol1_{a\in S}
       +\sum_{o\sim a,\,\pi(o)\ne a}\boldsymbol1_{\pi(o)\in S}\right).
 \label{eq:account-definition}
\end{align}
The first line charges the groups owned by $a$ after their residual
credits. The second line pays for the contributions of $a$ to groups
owned elsewhere. The third line reserves deletion costs. It counts
separate incidences even when several of them lead to the same group,
so it can reserve the cost of a deletion more than once.

Each incidence $ao$ is called a \emph{slot} at $a$. It is \emph{owned}
if $\pi(o)=a$, and \emph{off-owner} otherwise. Each current vertex has
three slots after padding. Table~\ref{tab:slot-charges} lists their
charges from the first two lines of \eqref{eq:account-definition}.
The ordinary or discounted status in the table always refers to the
owner of $o$.

\begin{table}[htbp]
\centering
\caption{The four slot charges, before deletion reserves.}
\label{tab:slot-charges}
\begin{tabular}{@{}l@{\qquad}l@{}}
\toprule
Slot at $a$ and owner status & Charge in $E_a$\\
\midrule
Owned, ordinary & $P_o-\tfrac12\sum_{b\sim o,\,b\ne a}t_b^r$\\[4pt]
Owned, discounted & $\tfrac56 P_o-\tfrac7{24}\sum_{b\sim o,\,b\ne a}t_b^r$\\[4pt]
Off-owner, ordinary & $J_{t_a}(F_{oa})+\tfrac12t_a^r$\\[4pt]
Off-owner, discounted & $\tfrac56J_{t_a}(F_{oa})+\tfrac7{24}t_a^r$\\
\bottomrule
\end{tabular}
\end{table}

\begin{lemma}\label{lem:account}
The account \eqref{eq:account-definition} satisfies $E_a\le3w_a/2$.
\end{lemma}

\subsection{Why the residual credit covers the extra weight}
We prove Lemma~\ref{lem:account} here for a discounted current vertex
that owns all three of its slots. This case shows how the residual
credit and the discount threshold work together.

Assume $t_a>0$, divide budgets by $t_a$ and weights by $t_a^r$, and use
the same notation for the normalized quantities. Then $t_a=1$.
Write $x_1,x_2,x_3$ for the three transfers and $z=\sum_i x_i$, so
$w_a=(1+z)^r$. Because $a$ is the owner, every outside residual is at
most $1$. If its two outside residual weights at slot $i$ are
$v_1,v_2\in[0,1]$, including zero values for missing neighbors, the slot
charge is
\[
 \tfrac56(x_i+v_1^{12/7}+v_2^{12/7})^{7/12}
       -\tfrac7{24}(v_1+v_2).
\]
The first term is a multiple of the $12/7$-norm of
$(x_i^{7/12},v_1,v_2)$. Hence the expression is convex in
$(v_1,v_2)$ and reaches its maximum on the square at a corner.
For an upper bound, we may therefore replace the outside weights by
zeros and ones. Let $j_i\in\{0,1,2\}$ count the ones at slot $i$ and
put $R=\sum_i j_i$.

All slots are owned, so the deletion reserve is $w_a/6$.
The remaining allowance for the three slot charges is $4w_a/3$.
H\"older's inequality gives
\[
 \begin{aligned}
 E_a-\tfrac16w_a
 &\le \tfrac56\sum_{i=1}^3(x_i+j_i)^r-\tfrac7{24}R\\
 &\le \tfrac56\,3^{5/12}(z+R)^r-\tfrac7{24}R
 \le \tfrac43(1+z+R)^r-\tfrac7{24}R,
 \end{aligned}
\]
where the last step uses $3^{5/12}\le8/5$.
The term $R$ is the extra budget supplied by outside residuals,
and $7R/24$ is the credit assigned to that budget. To compare with the
allowance $4(1+z)^r/3$, apply the tangent bound for the concave function
$u^r$ at $u=1+z$:
\[
 \tfrac43(1+z+R)^r-\tfrac7{24}R
 \le \tfrac43(1+z)^r+
 \left[\tfrac43 r(1+z)^{-5/12}-\tfrac7{24}\right]R.
\]
The expression in brackets is nonpositive exactly when
$1+z\ge(8/3)^{12/5}$. This is precisely the rule that makes a center
with three owned slots discounted. The credit thus covers the full
increase caused by $R$. Restoring the reserve gives
$E_a\le(4/3+1/6)w_a=3w_a/2$.

For a numerical example, take $x_i=5$ and $j_i=1$ for all three slots.
Then $1+z=16$ is above the threshold, and $R=3$ gives credit $7/8$.
The account is
\[
 E_a=\frac52 6^{7/12}-\frac78+\frac16 16^{7/12}
 \approx7.075,
 \qquad
 \frac32w_a=\frac32 16^{7/12}\approx7.560.
\]

The discounted case with two owned slots uses a bound with coefficient
$7/6$ in place of $4/3$. The same tangent calculation gives its threshold
$(7/3)^{12/5}$. Ordinary centers with two or three owned slots have
budgets below their corresponding thresholds, and are handled by
estimates on these bounded ranges. Centers with at most one owned slot
are treated directly. Appendix~\ref{sec:local} gives these remaining
scalar cases and the limiting argument for $t_a=0$.

The exponent $p=12/7$ makes the residual credits, H\"older bounds,
and threshold estimates hold together. Section~\ref{sec:scope}
discusses what the analysis establishes about other powers.

\subsection{Summing the local bounds}
We now finish the proof of Theorem~\ref{thm:mixed}. In $\sum_a E_a$,
each residual credit cancels the matching payment at the other current
vertex. By \eqref{eq:split}, the remaining main charges and increments
sum to at least $\sum_o s_{\pi(o)}w_o$.

For a discounted group $C_a$, the reserve in $E_a$ includes $w_a/6$.
Every other $b\in\N(C_a)$ is adjacent to some $o\in C_a$, and that
off-owner slot reserves $w_b/6$ in $E_b$. Thus the total reserve is at
least $\frac16\sum_{a\in S}w(\N(C_a))$, even when several incidences
reserve the same deletion. Lemma~\ref{lem:account} consequently gives
\[
 \sum_o s_{\pi(o)}w_o+\tfrac16\sum_{a\in S}w(\N(C_a))
 \le\sum_a E_a\le\tfrac32w(A).
\]
This is \eqref{eq:certificate}, which implies the desired inequality
\eqref{eq:theorem}.

\section{From local optimality to a weight bound}\label{sec:finite}
We now construct the transfers and residuals required by
Theorem~\ref{thm:mixed}. The starting point is that every small insertion
from the comparison packing has nonpositive potential gain. We express
these gains by a recursion on trees and use local optimality to bound
how much the recursion can assign to each current set. At a fixed depth,
a small increase in the current budgets makes the demand constraints
hold. The sum of these increases over all depths is bounded, which
leads to the following estimate.

\begin{theorem}\label{thm:finite}
Let $A,O$ be packings with $A\cap O=\varnothing$, let $v$ be
nonnegative weights with $v(o)>0$ for every $o\in O$, and let $H\ge1$
be an integer. Set $p=12/7$ and $r=7/12$. If
\begin{equation}
 \sum_{o\in X}v(o)^p\le\sum_{a\in\N_A(X)}v(a)^p
 \quad\text{for every }X\subseteq O\text{ with }|X|\le4^H-1,
 \label{eq:comparison-local-optimality}
\end{equation}
then
\begin{equation}
 v(O)\le\alpha_Hv(A)+\frac16\G_v(A,O),
 \qquad \alpha_H=\frac32+\frac7{8H}.
 \label{eq:finite-certificate}
\end{equation}
\end{theorem}

Throughout this section, write $c_a=v(a)^p$ and $q_o=v(o)^p$.
Thus $c_a$ is the potential lost by deleting $a$, and $q_o$ is the
potential gained by inserting $o$. Every $o\in O$ has a neighbor in
$A$, since otherwise $X=\{o\}$ violates
\eqref{eq:comparison-local-optimality}. The comparison graph has maximum
degree three on both sides. All the constructions below are used in the
proof to compare $A$ and $O$.

\subsection{Computation trees and projection}
The comparison graph may contain cycles, so exchanges explored from
different neighbors can meet again. We unfold the graph into a tree,
allowing separate copies of the same original vertex. Branches of this
tree can then be combined while keeping track of every cost.

Fix a root in a connected component. A vertex of the unfolded tree is a
walk from the root that never immediately reverses an edge. Its label is
the last vertex of the walk. A walk is adjacent to its predecessor and
to each allowed one-edge extension. Consequently, each copy has exactly
one neighbor of every label adjacent to its own label in the original
graph. In particular, the tree edges corresponding to any fixed
original edge $oa$ form a matching.

A tree exchange selects finitely many comparison copies and deletes
all their neighboring current copies. Its gain is the sum of their
inherited demands $q_o$ minus the sum of the inherited costs $c_a$ in
this full neighborhood. Each copy is counted once. To obtain an
insertion in the original graph, we group copies according to how often
each label occurs.

\begin{lemma}\label{lem:projection}
For any integer $b\ge1$, a positive-gain tree exchange containing at
most $b$ comparison copies yields an insertion $X\subseteq O$ with
positive potential gain and $|X|\le b$.
\end{lemma}
\begin{proof}
Let $m_o$ count the selected copies of $o$, and let $n_a$ count the copies
of $a$ in their full neighborhood. On each original edge $oa$, the
matching property gives a distinct neighboring copy of $a$ for every
selected copy of $o$. Hence $n_a\ge m_o$.

For each integer $j\ge1$, put
\[
 I_j=\{o\in O:m_o\ge j\},\qquad
 D_j=\{a\in A:n_a\ge j\}.
\]
These are the labels appearing at least $j$ times on the two sides.
The preceding inequality implies $\N_A(I_j)\subseteq D_j$. Counting
each label once for every copy gives
\[
 \begin{aligned}
 \sum_{j\ge1}\left(\sum_{o\in I_j}q_o-
                  \sum_{a\in\N_A(I_j)}c_a\right)
 &\ge \sum_{j\ge1}\left(\sum_{o\in I_j}q_o-
                        \sum_{a\in D_j}c_a\right)\\
 &=\sum_o m_oq_o-\sum_a n_ac_a>0.
 \end{aligned}
\]
All sums are finite, so some $I_j$ has positive gain. It is a packing
because $I_j\subseteq O$, and its size is at most
$\sum_o m_o\le b$.
\end{proof}

\subsection{Branch gains and current budgets}
Consider a comparison copy labeled $o$ with a designated current parent
labeled $a$. We evaluate a branch below $o$ assuming that the cost of
$a$ has already been paid. Selecting $o$ earns $q_o$ and requires us to
pay for its other current neighbors. Further comparison branches can
then be attached through these paid neighbors. A nonempty branch
contains $o$, and the empty branch has gain zero.

For an edge $oa$, let the nonnegative number $x^h_{oa}$ describe such a
conditional gain with at most $h$ comparison layers. We call these
numbers messages. Their recursion is
\begin{equation}
 x^0_{oa}=0,\qquad
 x^{h+1}_{oa}=\left[q_o-
 \sum_{b\in\N(o)\setminus\{a\}}
 \left(c_b-\sum_{z\in\N(b)\setminus\{o\}}x^h_{zb}\right)\right]_+.
 \label{eq:finite-messages}
\end{equation}
At depth one, this is the gain of inserting $o$ alone with the parent
cost excluded, or zero if that gain is negative. At larger depths,
each inner sum is the gain available from branches attached through
$b$. The cost $c_b$ is paid once for all of them. The positive part
again allows the empty branch.

\begin{figure}[t]
\centering
\begin{minipage}{.42\textwidth}
\centering
\begin{tikzpicture}[line width=.5pt,
 current/.style={draw,rectangle,minimum size=5mm,inner sep=.8pt},
 comparison/.style={draw,circle,minimum size=5mm,inner sep=.8pt}]
 \node[current] (a) at (0,0) {$a$};
 \node[comparison] (o) at (0,-.75) {$o$};
 \node[current] (b) at (-1.1,-1.5) {$b$};
 \node[current] (c) at (1.1,-1.5) {$c$};
 \node[comparison] (z1) at (-1.65,-2.3) {$z_1$};
 \node[comparison] (z2) at (-.55,-2.3) {$z_2$};
 \node[comparison] (z3) at (.55,-2.3) {$z_3$};
 \node[comparison] (z4) at (1.65,-2.3) {$z_4$};
 \draw[dashed] (a)--(o);
 \draw (o)--(b) (o)--(c) (b)--(z1) (b)--(z2) (c)--(z3) (c)--(z4);
 \draw[dashed] (z1)--++(0,-.3) (z2)--++(0,-.3)
               (z3)--++(0,-.3) (z4)--++(0,-.3);
\end{tikzpicture}
\end{minipage}\hfill
\begin{minipage}{.54\textwidth}
\caption{One step of the branch recursion. The parent cost $c_a$ has
already been paid. The branch pays $c_b$ and $c_c$ once each. The
branches rooted at $z_1,z_2$ exclude $c_b$, and those rooted at $z_3,z_4$
exclude $c_c$. Their remaining subtrees are disjoint. Labels may repeat
elsewhere in the unfolded tree.}
\label{fig:computation-branch}
\end{minipage}
\end{figure}
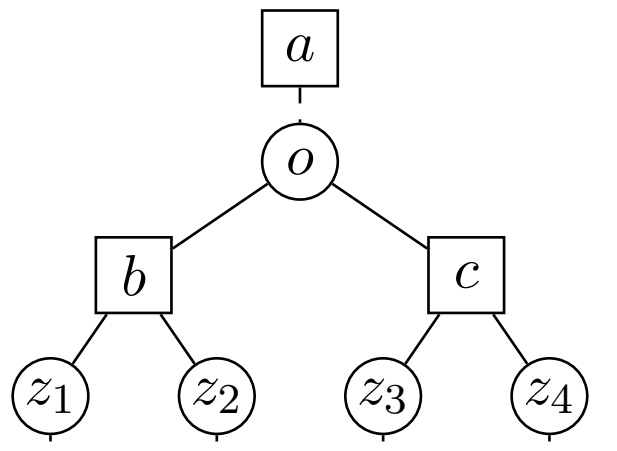

The next lemma verifies that each message is realized by an actual tree
branch. It then bounds the total incoming gain at a current vertex.
If this gain exceeded its cost, joining the branches and paying that
cost would give a forbidden improvement.

\begin{lemma}\label{lem:message-capacity}
Each $x^h_{oa}$ is the conditional gain of a branch with at most
$T_h=(4^h-1)/3$ comparison copies. The messages are nondecreasing in $h$,
and under \eqref{eq:comparison-local-optimality},
\begin{equation}
 \ell_a^h:=\sum_{o\sim a}x^h_{oa}\le c_a
 \qquad(a\in A,\ 0\le h\le H).
 \label{eq:capacity}
\end{equation}
\end{lemma}
\begin{proof}
At depth zero use the empty branch. Suppose the claim holds at depth
$h$. When the bracket in \eqref{eq:finite-messages} is positive, select
$o$, pay for its nonparent neighbors $b$, and attach their positive
depth-$h$ descendant branches. Each attached branch excludes the
already-paid cost of its parent $b$. Its remaining vertices lie in a
separate subtree, so the gains add to the displayed bracket. When the
bracket is nonpositive, use the empty branch. Thus every selected
comparison copy has its full current neighborhood paid, except for
the designated parent of the whole branch.

There are at most two nonparent neighbors $b$, each with at most two
further comparison neighbors. Therefore
\[
 T_0=0,\qquad T_{h+1}=1+4T_h,\qquad T_h=(4^h-1)/3.
\]
The recursion is nondecreasing in every message on its right-hand side.
Starting from zero gives $0\le x^h_{oa}\le x^{h+1}_{oa}$ by induction.

Now root the unfolded tree at a current copy $a$ and attach its at most
three depth-$h$ branches. If $\ell_a^h>c_a$, their union has positive
gain after the root cost is paid. Its full neighborhood is paid, and
it contains at most $3T_h=4^h-1\le4^H-1$ comparison copies.
Lemma~\ref{lem:projection} would give a positive-gain insertion in the
original graph, contradicting \eqref{eq:comparison-local-optimality}.
\end{proof}

\subsection{Completing the demand constraints}
At depth $h$, the choice $F_{oa}=x^h_{oa}$ leaves residual
$t_a^h=c_a-\ell_a^h\ge0$ at each current vertex. It satisfies the
budget equalities of Theorem~\ref{thm:mixed}, but a comparison demand
may still exceed its transfers and available residuals. We meet this
demand by taking one further message increment on an edge to a neighbor
with largest residual. This is the neighbor whose residual the demand
constraint omits. The added transfers require additional budgets
$d_a^h$. Their total over the $H$ depths is at most the original budget
$c_a$.

\begin{lemma}\label{lem:capacity-completion}
For every $0\le h<H$, there are nonnegative residuals $t_a^h$,
transfers $F_{oa}^h$, and budget increases $d_a^h$ such that
\begin{align*}
 c_a+d_a^h&=t_a^h+\sum_{o\sim a}F_{oa}^h,\\
 q_o&\le\sum_{a\sim o}F_{oa}^h+
          \sum_{a\sim o}t_a^h-\max_{a\sim o}t_a^h,
\end{align*}
and
\begin{equation}
 d_a^h\le\ell_a^{h+1}-\ell_a^h,
 \qquad \sum_{h=0}^{H-1}d_a^h\le c_a.
 \label{eq:telescope}
\end{equation}
\end{lemma}
\begin{proof}
Set $t_a^h=c_a-\ell_a^h$. For each $o$, choose
$a_h(o)\in\N(o)$ with largest residual $t_a^h$, and define
\[
 u_o^h=x^{h+1}_{o,a_h(o)}-x^h_{o,a_h(o)},\qquad
 F_{oa}^h=x^h_{oa}+\boldsymbol1\{a=a_h(o)\}u_o^h,
 \qquad d_a^h=\sum_{o:a_h(o)=a}u_o^h.
\]
All these quantities are nonnegative by
Lemma~\ref{lem:message-capacity}. The budget equality follows directly
from the definitions. To check the demand constraint, note that
\[
 c_b-\sum_{z\in\N(b)\setminus\{o\}}x^h_{zb}=t_b^h+x^h_{ob}.
\]
The recurrence \eqref{eq:finite-messages} therefore gives, for every
$a\sim o$,
\[
 q_o\le\sum_{b\in\N(o)\setminus\{a\}}(t_b^h+x^h_{ob})
                    +x^{h+1}_{oa}.
\]
Choose $a=a_h(o)$. The residual sum omits a largest residual, and the
transfer sum includes the increment $u_o^h$. Hence
\[
 q_o\le\sum_{a\sim o}F_{oa}^h+
          \sum_{a\sim o}t_a^h-\max_{a\sim o}t_a^h.
\]

For fixed $a$, $d_a^h$ sums a subset of the nonnegative increments
on the edges incident to $a$. Thus
\[
 d_a^h\le\sum_{o\sim a}(x^{h+1}_{oa}-x^h_{oa})
             =\ell_a^{h+1}-\ell_a^h.
\]
Summing over depths cancels the intermediate loads. Since
$\ell_a^0=0$ and $\ell_a^H\le c_a$, this proves
\eqref{eq:telescope}.
\end{proof}

\subsection{Averaging over depths}
The completed decompositions allow us to apply
Theorem~\ref{thm:mixed} at each depth. Concavity converts their bounded
total budget increase into an average weight increase of order $1/H$.

\begin{proof}[Proof of Theorem~\ref{thm:finite}]
At depth $h$, apply Theorem~\ref{thm:mixed} with current budgets
$c_a+d_a^h$ and demands $q_o$. The comparison weights are
$q_o^r=v(o)$, and the current weights are $(c_a+d_a^h)^r$.
Increasing a current weight can only reduce every group's gain,
because that weight contributes to its deletion cost. The resulting
maximum assignment gain is therefore at most $\G_v(A,O)$. We obtain
\[
 v(O)\le\frac32\sum_a(c_a+d_a^h)^r+\frac16\G_v(A,O)
 \qquad(0\le h<H).
\]
For $c_a>0$, the tangent bound for the concave function $y^r$ gives
\[
 \begin{aligned}
 \frac1H\sum_{h=0}^{H-1}(c_a+d_a^h)^r
 &\le c_a^r+\frac rHc_a^{r-1}\sum_{h=0}^{H-1}d_a^h\\
 &\le\left(1+\frac rH\right)c_a^r,
 \end{aligned}
\]
where the second inequality uses \eqref{eq:telescope}. If $c_a=0$,
that same bound forces $d_a^h=0$ at every depth, so the conclusion
still holds. Average the preceding inequalities over $h$ and use
$c_a^r=v(a)$ and $(3/2)r=7/8$. This gives
\[
 v(O)\le\frac32\left(1+\frac rH\right)v(A)+\frac16\G_v(A,O)
       =\alpha_Hv(A)+\frac16\G_v(A,O),
\]
as required.
\end{proof}

\section{Approximation guarantee and running time}\label{sec:rounding}
\label{sec:implementation-details}

We now connect the structural bound to the algorithm's original
weights. The first phase compares integer values $Q$, the structural
analysis uses power weights, and the second phase uses $w$.
Auxiliary weights $v$ link these three views. We first bound the loss
from rounding, then compare with an optimum after removing common sets.

\subsection{Rounding and the assignment gain}
Recall the auxiliary weights $v(S)=\tau Q(S)^r$ introduced
in Section~\ref{sec:algorithm}, where $p=12/7$ and $r=7/12$. Since $Q(S)\le(w(S)/\tau)^p<Q(S)+1$ and $(Q+1)^r-Q^r\le1$ for every $Q\ge0$, we have
\begin{equation}
 0\le w(S)-v(S)\le\tau.
 \label{eq:rounding}
\end{equation}
The identity $v(S)^p=\tau^pQ(S)$ makes the first phase an exact
power-potential local search for $v$. Sets with $v(S)=0$ may be omitted
from this phase: removing them from an insertion leaves its potential
contribution of the inserted sets unchanged and can only reduce its deletion cost. Thus a positive
improvement involving such sets would imply a positive improvement
using only positive-$v$ sets. All original sets remain available in
the second phase.

\begin{lemma}\label{lem:rounded-gain}
For every packing $O$ disjoint as a family from $A$, all of whose
members have a neighbor in $A$,
\[
 \G_v(A,O)\le9\Delta+3n\tau.
\]
\end{lemma}
\begin{proof}
For any assignment group $C$,
\[
 [v(C)-v(\N_A(C))]_+
 \le[g_A(C)]_++\sum_{a\in\N_A(C)}(w(a)-v(a)).
\]
A current set occurs in at most three group deletion neighborhoods,
because it has at most three neighbors in $O$. Summing over all groups,
using \eqref{eq:rounding}, and applying Lemma~\ref{lem:greedy} gives the
claim for each assignment, hence for their maximum.
\end{proof}

\subsection{Common-member cancellation and the approximation guarantee}
Let $O^\star$ be an optimum packing on the preprocessed instance and set
\[
 C=A\cap O^\star,\qquad A_0=A\setminus C,\qquad
 O_0=\{o\in O^\star\setminus C:v(o)>0\}.
\]
Every $o\in O_0$ has a neighbor in $A$ by single-set local optimality.
No such neighbor belongs to $C$, since $O^\star$ is a packing.
Consequently, for every $X\subseteq O_0$,
\begin{equation}
 \N_{A_0}(X)=\N_A(X),\qquad
 \G_v(A_0,O_0)=\G_v(A,O_0).
 \label{eq:common-cancellation}
\end{equation}
For the second equality, observe that assignments, nonempty groups, and
their full deletion neighborhoods agree on the two comparison graphs.
The discarded vertices of $A$ have no neighbors in $O_0$.

The first phase is locally optimal for $v$. For every $X\subseteq O_0$
with $|X|\le4^H-1$, \eqref{eq:common-cancellation} therefore gives
$\Phi_p^v(X)\le\Phi_p^v(\N_{A_0}(X))$. These are the hypotheses of
Theorem~\ref{thm:finite} for $(A_0,O_0)$. Restoring the common
contribution yields
\[
 \begin{aligned}
 v(O^\star)
 &=v(C)+v(O_0)\\
 &\le v(C)+\alpha_Hv(A_0)+\tfrac16\G_v(A_0,O_0)\\
 &\le\alpha_Hv(A)+\tfrac16\G_v(A,O_0),
 \end{aligned}
\]
where $\alpha_H\ge1$ and discarded zero-$v$ members contribute nothing.
This step does not require $A_0$ to be locally optimal on the full input.
Lemma~\ref{lem:rounded-gain}, \eqref{eq:rounding}, and
$w(B)\ge w(A)+\Delta$ now imply
\[
 \begin{aligned}
 w(O^\star)
 &\le v(O^\star)+n\tau\\
 &\le\alpha_Hv(A)+\tfrac16\G_v(A,O_0)+n\tau\\
 &\le\alpha_Hw(A)+\tfrac32\Delta+\tfrac32n\tau\\
 &\le\alpha_Hw(B)+\tfrac32n\tau.
 \end{aligned}
\]
Since $M\le w(O^\star)$, the parameter choices give
\[
 \left(1-\frac\eps{20}\right)w(O^\star)
 \le\left(\frac32+\frac{7\eps}{8}\right)w(B).
\]
For $0<\eps\le1$, the exact identity
\[
 \left(\frac32+\eps\right)\left(1-\frac\eps{20}\right)
 -\left(\frac32+\frac{7\eps}{8}\right)
 =\frac{\eps(1-\eps)}{20}\ge0
\]
proves the desired ratio. Adding back the packing $P$ of positive-weight
empty sets preserves it, because the ratio is at least one. Zero-weight
sets may be discarded. If no sets remain after preprocessing, returning
$P$ is optimal. For $\eps>1$, running the algorithm with accuracy one
gives factor $5/2\le3/2+\eps$.

\subsection{Termination and bit complexity}
Every improving move increases the integer potential by at least one,
and
\[
 Q(S)\le(30n/\eps)^{12/7},\qquad
 Q(A)\le n(30n/\eps)^{12/7}.
\]
Thus there are at most $n(30n/\eps)^{12/7}$ first-phase iterations.
An exhaustive scan of insertions of size at most $b$ takes
$n^{\OO(b)}\operatorname{poly}(L)$ time for fixed $\eps$. Testing
disjointness, constructing deletion neighborhoods, and summing integer
contributions take polynomial time per candidate. The second phase
enumerates $\OO(n^3)$ candidates, sorts their rational gains, and tests
compatibility in polynomial time. Since
$b=4^{\lceil1/\eps\rceil}-1$, the total running time is
$n^{4^{\OO(1/\eps)}}\operatorname{poly}(L)$.

The rounded contributions require no comparison between irrational
powers. Write $w(S)/\tau=u/d$ as a positive rational number. For an
integer $z\ge0$,
\[
 z\le(u/d)^{12/7}\quad\Longleftrightarrow\quad
 z^7d^{12}\le u^{12}.
\]
Binary search on $z$ computes $Q(S)$ with integer arithmetic. The
integers have polynomial encoding length, and the second-phase rational
operations also have polynomial bit complexity. The virtual weights
$v$ are never computed. This proves the running-time bound in Theorem~\ref{thm:algorithm}.
The graph extension is proved next.

\section{Extensions and limits}\label{sec:scope}
The same analysis applies to a larger graph class. We then discuss
what the proof says about the choice of exponent and about the limits
of the two-phase scheme.

\subsection{Independent sets in \texorpdfstring{induced $K_{1,4}$-free graphs}{induced K1,4-free graphs}}
\label{sec:graph-extension-proof}

A graph is \emph{induced $K_{1,4}$-free} if no vertex has four
pairwise nonadjacent neighbors. Weighted $3$-set packing is a special
case of maximum-weight independent set in such graphs: make one vertex
for each input set and join two vertices when their sets intersect.
Four pairwise disjoint sets cannot all meet one set of size at most three.
The next proof shows that the algorithm and analysis use only the
resulting bound on conflicts with an independent set.

\begin{proof}[Proof of the graph extension in Theorem~\ref{thm:algorithm}]
It suffices to take $0<\eps\le1$; for larger $\eps$, run with
accuracy one. Replace packings by independent sets, and define $\N_A(X)$ as the
neighbors of $X$ in the current independent set $A$. Discard zero-weight
vertices and handle an empty remaining graph directly. Apply the same
integer rounding and first-phase search. In the second phase, enumerate
independent insertions $X$ of at most three vertices with positive gain.
If $U$ is the set inserted so far, accept $X$ exactly when
$X\cap N_G[U]=\varnothing$, where $N_G[U]$ is the closed neighborhood.
This ensures both independence and absence of repeated inserted vertices.

For two disjoint independent sets $A,O$, each vertex has at most three
neighbors on the other side; otherwise those neighbors and the vertex
induce a $K_{1,4}$. Thus Theorem~\ref{thm:mixed} applies to their
comparison graph. Lemma~\ref{lem:projection} applies because every subset of $O$ is independent
and the lifts of each comparison edge form a matching. The degree bound
of three gives the branch-size estimate in Lemma~\ref{lem:message-capacity}.
The proof of Lemma~\ref{lem:capacity-completion} and the subsequent
averaging argument use the same budgets and messages. Therefore
Theorem~\ref{thm:finite} holds for these independent sets as well.

It remains to establish the recovery factor in the graph model. For
any independent comparison set $O$ and candidate $X$ with $|X|\le3$,
\[
 |N_G[X]\cap O|\le3|X|\le9.
\]
Indeed, if $x\in O$, then $N_G[x]\cap O=\{x\}$; if $x\notin O$,
it has at most three neighbors in $O$. The comparison groups are
disjoint subsets of $O$. Hence an accepted candidate blocks at most
nine groups, with blocking understood to include a common vertex.
The first-blocker charging argument in Lemma~\ref{lem:greedy} gives
$\G_w\le9\Delta$. The accepted insertions are disjoint and independent,
so $w(B)-w(A)\ge\Delta$ still follows by counting every deletion once.

Finally, a current vertex lies in at most three group deletion
neighborhoods, again by the comparison degree bound. Lemma~\ref{lem:rounded-gain}
and the approximation proof therefore apply. Common-vertex cancellation
is valid because the optimum is independent. Integer-potential termination,
exact comparisons, and candidate enumeration have the same bounds.
\end{proof}

\subsection{The exponent and the two-phase scheme}
The exponent $12/7$ is chosen so that the residual credits, H\"older
bounds, and discount thresholds in Section~\ref{sec:mixed-proof}
work together. It is sufficient for this proof; we do not claim that
it is the only possible choice. Appendix~\ref{sec:exponent-examples}
shows that replacing it by a fixed exponent
\[
 0<p<\log_{3/2}2\qquad\text{or}\qquad p\ge1.83
\]
can make the approximation ratio strictly worse than $3/2$, even after
the same recovery phase. These examples do not establish that every
exponent between the two bounds works. Our analysis does not determine
whether other exponents yield the same guarantee.

A separate construction in Appendix~\ref{sec:framework-lower-bound}
applies to every power $p>0$ and every fixed first-phase insertion
bound. It gives a unit-weight instance and a permitted first-phase
execution whose output has ratio exactly $3/2$ and admits no
positive-gain recovery candidate. Thus a guarantee valid for all
permitted first-phase choices cannot be better than $3/2$ in this
framework. This is a limitation of the scheme under allowed choices,
not a hardness result for the problem. It also leaves room for an
exact $3/2$ guarantee or for an algorithm that imposes further
conditions on the first-phase choices.

The current algorithm uses a large neighborhood to make the depth
error small. Reducing this neighborhood, or obtaining an exact
polynomial-time $3/2$ guarantee, are natural directions for further
work.

\label{maintext:end}
\paragraph{Acknowledgements.}
Generative AI tools, including ChatGPT and Codex, were used to assist
with exploratory mathematical reasoning, proof checking, and the
drafting and organization of this manuscript. The authors take full
responsibility for the correctness of the mathematical claims and
arguments, the accuracy of the references, and the originality and
final presentation of the work.

\bibliographystyle{plainnat}
\bibliography{references}

@inproceedings{berman2000,
  author    = {Piotr Berman},
  title     = {A $d/2$ Approximation for Maximum Weight Independent Set in $d$-Claw Free Graphs},
  booktitle = {Algorithm Theory---SWAT 2000},
  series    = {Lecture Notes in Computer Science},
  volume    = {1851},
  pages     = {214--219},
  publisher = {Springer},
  year      = {2000},
}

@inproceedings{thieryward2023,
  author    = {Theophile Thiery and Justin Ward},
  title     = {An Improved Approximation for Maximum Weighted $k$-Set Packing},
  booktitle = {Proceedings of the 2023 Annual ACM-SIAM Symposium on Discrete Algorithms (SODA)},
  pages     = {1138--1162},
  publisher = {SIAM},
  year      = {2023},
}

@article{neuwohner2024,
  author    = {Meike Neuwohner},
  title     = {The limits of local search for weighted $k$-set packing},
  journal   = {Mathematical Programming},
  volume    = {206},
  pages     = {389--427},
  year      = {2024},
}

@inproceedings{fureryu2014,
  author    = {Martin F{\"u}rer and Huiwen Yu},
  title     = {Approximating the $k$-Set Packing Problem by Local Improvements},
  booktitle = {Combinatorial Optimization (ISCO 2014)},
  series    = {Lecture Notes in Computer Science},
  volume    = {8596},
  pages     = {408--420},
  publisher = {Springer},
  year      = {2014},
}

@misc{tongxu2026,
  author        = {Weitian Tong and Yao Xu},
  title         = {Breaking the $\sqrt{3}$ Barrier for Maximum Weighted $3$-Set Packing},
  year          = {2026},
  eprint        = {2610.10036},
  archivePrefix = {arXiv},
  primaryClass  = {cs.DS},
  note          = {Preprint, version 1, 7 October 2026},
  url           = {https://arxiv.org/html/2610.10036v1}
}

@article{AH98,
  author  = {Esther M. Arkin and Refael Hassin},
  title   = {On local search for weighted {k}-set packing},
  journal = {Mathematics of Operations Research},
  volume  = {23},
  number  = {3},
  pages   = {640--648},
  year    = {1998}
}

@article{BMR09,
  author  = {P{\'e}ter Bir{\'o} and David F. Manlove and Romeo Rizzi},
  title   = {Maximum weight cycle packing in directed graphs, with application to kidney exchange programs},
  journal = {Discrete Mathematics, Algorithms and Applications},
  volume  = {1},
  number  = {4},
  pages   = {499--517},
  year    = {2009}
}

@article{CC06,
  author  = {Miroslav Chleb{\'\i}k and Janka Chleb{\'\i}kov{\'a}},
  title   = {Complexity of approximating bounded variants of optimization problems},
  journal = {Theoretical Computer Science},
  volume  = {354},
  number  = {3},
  pages   = {320--338},
  year    = {2006}
}

@article{CH01,
  author  = {Barun Chandra and Magn{\'u}s M. Halld{\'o}rsson},
  title   = {Greedy local improvement and weighted set packing approximation},
  journal = {Journal of Algorithms},
  volume  = {39},
  number  = {2},
  pages   = {223--240},
  year    = {2001}
}

@inproceedings{Cyg13,
  author    = {Marek Cygan},
  title     = {Improved approximation for {3}-dimensional matching via bounded pathwidth local search},
  booktitle = {FOCS},
  pages     = {509--518},
  year      = {2013}
}

@article{HS89,
  author  = {Cor A. J. Hurkens and Alexander Schrijver},
  title   = {On the size of systems of sets every {t} of which have an {SDR}, with an application to the worst-case ratio of heuristics for packing problems},
  journal = {SIAM Journal on Discrete Mathematics},
  volume  = {2},
  number  = {1},
  pages   = {68--72},
  year    = {1989}
}

@article{HSS06,
  author  = {Elad Hazan and Shmuel Safra and Oded Schwartz},
  title   = {On the complexity of approximating {k}-set packing},
  journal = {Computational Complexity},
  volume  = {15},
  number  = {1},
  pages   = {20--39},
  year    = {2006}
}

@inproceedings{LST25,
  author    = {Euiwoong Lee and Ola Svensson and Theophile Thiery},
  title     = {Asymptotically optimal hardness for $k$-set packing and $k$-matroid intersection},
  booktitle = {STOC},
  pages     = {54--61},
  year      = {2025}
}

@inproceedings{Neu21,
  author    = {Meike Neuwohner},
  title     = {An improved approximation algorithm for the maximum weight independent set problem in $d$-claw free graphs},
  booktitle = {STACS},
  volume    = {187},
  pages     = {53:1--53:20},
  year      = {2021}
}

@inproceedings{Neu23,
  author    = {Meike Neuwohner},
  title     = {Passing the limits of pure local search for weighted $k$-set packing},
  booktitle = {SODA},
  pages     = {1090--1137},
  year      = {2023}
}

@misc{ST26,
  author = {Neta Singer and Theophile Thiery},
  title  = {Submodular maximization over many matroids via ordered local search},
  year   = {2026},
  note   = {arXiv preprint}
}

@phdthesis{thiery2023thesis,
  author = {Theophile Fran{\c{c}}ois Thiery},
  title  = {Approximation Algorithms for Independence Systems},
  school = {Queen Mary University of London},
  month  = {August},
  year   = {2023}
}
\clearpage
\appendix
\renewcommand{\theHsection}{appendix.\Alph{section}}
\renewcommand{\theHsubsection}{\theHsection.\arabic{subsection}}

\section{Complete scalar verification of the account bound}\label{sec:local}
We prove Lemma~\ref{lem:account}, the bound $E_a\le 3w_a/2$ used in the proof of Theorem~\ref{thm:mixed}. We first normalize the local parameters,
then verify the discounted-center and ordinary-center cases, and
finally handle zero residuals.

\subsection{Normalization and local parameters}

First assume \(t_a>0\). Measure budgets in units of \(t_a\), and
weights in units of \(t_a^r\). The residual at the vertex under
inspection is now \(1\).

Number its three incident slots so that the owned comparison
vertices come first. Let their normalized transfers be
\(x_1,x_2,x_3\ge0\), and write
\[
z=x_1+x_2+x_3.
\]
The normalized current budget is \(1+z\), and its weight is
\((1+z)^r\).

Only the following local counts will be used. In this appendix, $k$
counts owned slots; it is not an input set-size parameter.
\begin{center}
\begin{tabular}{@{}lp{0.73\textwidth}@{}}
\toprule
Symbol&Meaning inside this section\\
\midrule
\(k\)&Number of owned slots, from \(0\) to \(3\).\\
\(m\)&Number of the other \(3-k\) slots whose owners are discounted.\\
\(j_i\)&Sum of outside residuals for owned slot \(i\), reduced below to \(0,1,2\).\\
\(R=\sum_{i=1}^k j_i\)&Total outside residuals  in these owned slots.\\
\bottomrule
\end{tabular}
\end{center}

Here is why outside residuals can be reduced to unit counts. For an owned comparison vertex, the other two residuals are at most \(1\), because its owner has maximum residual. If their residual weights (the
$7/12$-th powers of their residuals) are \(v_1,v_2\in[0,1]\), the charge minus residual credit from this slot has the form
\[
s_a\cdot
(x_i+v_1^{12/7}+v_2^{12/7})^{7/12}
-\lambda_a\cdot(v_1+v_2).
\]
The first term is a multiple of the \(12/7\)-norm of
\((x_i^{7/12},v_1,v_2)\), so it is convex in \(v_1,v_2\).
Subtracting a linear function preserves convexity.
A maximum on the square \([0,1]^2\) occurs at a corner.
Each residual can therefore be replaced, for an upper bound, by \(0\) or \(1\). Their sum is \(j_i\).

Table~\ref{tab:slot-charges} gives the four account entries in the
original variables. After normalization and the corner reduction, an
owned entry becomes $s_a(x_i+j_i)^r-\lambda_a j_i$, whereas an
off-owner entry becomes $s_{\pi(o)}J_1(x_i)+\lambda_{\pi(o)}$.
The deletion reserve takes \(m/6\) of the current weight, plus
another \(1/6\) if the current center itself is discounted.
Therefore the remaining multiplier is
\begin{equation}
\Lambda=
\begin{cases}
3/2-m/6,&\text{current center ordinary},\\
4/3-m/6,&\text{current center discounted}.
\end{cases}
\label{eq:localbudget}
\end{equation}
The precise local goal is
\begin{equation}
\text{sum of the three slot charges}\le \Lambda(1+z)^r.
\label{eq:localgoal}
\end{equation}
This is Lemma~\ref{lem:account} after the rescaling.
These terms are the normalized entries of \eqref{eq:account-definition}.

\subsection{H\"older and tangent inequalities}

We use H\"older in the explicit form
\begin{equation}
\sum_i b_i y_i^r
\le
\left(\sum_i b_i^{12/5}\right)^{5/12}
\left(\sum_i y_i\right)^r.
\label{eq:holder}
\end{equation}
It combines several charges into one charge depending on their total budget. Its coefficient is the \(12/5\)-norm of the numerical coefficients \(b_i\).
We also use the tangent bound
\[
(X+Y)^r\le X^r+rX^{-5/12}Y\qquad(X>0,\ Y\ge0).
\]
It controls the extra charge caused by adding residual mass.

\subsection{Discounted centers}
\label{sec:discounted}
By definition, a discounted center owns either two or three vertices.
First reduce the two-owned-slot case to a discounted third owner.
An ordinary third owner is easier:
Moving that owner's possible deletion reserve to the left,
changing it from ordinary to discounted increases the required
left side by
\[
\frac{(1+z)^r-J_1(x_3)}6-\frac5{24}\ge0.
\]
By the monotonicity of $J_t(x)$ in $t$
(\Appendixref{sec:mechanical}), we have
$J_1(x)\le J_0(x)=(2^r-1)x^r$, so
\(J_1(x)\le(2^r-1)x^r\le(1+z)^r/2\);
the discount threshold gives \((1+z)^r>5/2\).
For the latter bound, use
\((7/3)^{7/5}>5/2\).
Thus it suffices to prove the case with a discounted third owner.

Only two configurations remain: two owned slots with a discounted
third owner, where \(\Lambda=7/6\), and three owned slots, where \(\Lambda=4/3\).
For these configurations, we will show that the slot charges are at most
\begin{equation}
\Lambda(1+z+R)^r-\frac7{24}R.
\label{eq:active-envelope}
\end{equation}

The subtraction is precisely the corresponding residual credit.
Once \eqref{eq:active-envelope} is obtained, the tangent bound gives
\[
\Lambda(1+z+R)^r
\le \Lambda(1+z)^r+\Lambda r(1+z)^{-5/12}R.
\]
In both configurations, the discount threshold gives
$1+z\ge(2\Lambda)^{12/5}$, and hence
\[
\Lambda r(1+z)^{-5/12}\le r/2=7/24.
\]
Thus the residual credit pays for the entire extra charge.
This proves \eqref{eq:localgoal}.

It remains to justify the envelope \eqref{eq:active-envelope}.

\paragraph{Three owned slots.}
Here \(\Lambda=4/3\). H\"older gives
\[
\frac56\sum_{i=1}^3(x_i+j_i)^r-\frac7{24}R
\le \frac56\,3^{5/12}(z+R)^r-\frac7{24}R
\le \frac43(1+z+R)^r-\frac7{24}R.
\]
The last coefficient bound follows from \(3^{5/12}\le8/5\).
Section~\ref{sec:mixed-proof} gives the full account calculation for this case.

\paragraph{Two owned slots.}
Now the third slot has a discounted owner, hence \(\Lambda=7/6\).
The useful off-owner bound is
\begin{equation}
\frac56J_1(x)+\frac7{24}
\le\frac56\cdot\frac{11}{20}(1+x)^r.
\label{eq:sharp-off}
\end{equation}
Here is the proof. After dividing by \(5/6\), the inequality says
\[
(1+2x)^r+\frac7{20}\le\frac{31}{20}(1+x)^r.
\]
Apply \eqref{eq:holder} to \(1+2x\) and \(1\).
It suffices that
\[
2^r\left(1+(7/20)^{12/5}\right)^{5/12}\le31/20.
\]
For an exact certificate, use
\[
2^r\le\frac32,\qquad
\left(\frac7{20}\right)^{12/5}\le\frac{81}{1000},\qquad
\left(\frac{1081}{1000}\right)^{5/12}\le\frac{31}{30}.
\]
The resulting bound is $(3/2)(31/30)=31/20$.
The rational-power comparisons used here and below are recorded in
Appendix~\ref{sec:constants}.

Using \cref{eq:sharp-off} and H\"older,
\[
\begin{split}
&\frac56\sum_{i=1}^2(x_i+j_i)^r
+\frac56J_1(x_3)+\frac7{24}-\frac7{24}R\\
&\qquad\le
\frac56\left(2+(11/20)^{12/5}\right)^{5/12}
            (1+z+R)^r-\frac7{24}R\\
&\qquad\le\frac76(1+z+R)^r-\frac7{24}R.
\end{split}
\]
The last coefficient bound follows from
\(
(11/20)^{12/5}\le6/25,\text{and }
(7/5)^{12/5}\ge56/25.
\)

We have finished all discounted centers.

\subsection{Ordinary centers}
\label{sec:ordinary}

The current center is now ordinary.
An owned slot never needs two unit residuals in the upper-bound
calculation: replacing \(j_i=2\) by \(j_i=1\) cannot lower its charge,
because
\[
(x_i+2)^r-(x_i+1)^r\le2^r-1\le1/2.
\]
Hence every \(j_i\) is now \(0\) or \(1\), and \(0\le R\le k\).

The useful off-owner estimates in this case are
\begin{equation}
\begin{array}{rcl}
J_1(x)+1/2&\le&\frac12(1+2x)^r,\\[2pt]
\frac56J_1(x)+\frac7{24}&\le&\frac13(1+2x)^r.
\end{array}
\label{eq:ordinary-off}
\end{equation}
The first is midpoint concavity.
For the second, rearrange it as
\[
\frac12(1+2x)^r+\frac7{24}\le\frac56(1+x)^r
\]
and apply \eqref{eq:holder}.
The required coefficient comparison follows from
\[
\frac12+(2^r\cdot\tfrac7{24})^{12/5}
\le\frac12+(7/16)^{12/5}
\le\frac12+\frac7{50}=\frac{16}{25}
\le(5/6)^{12/5}.
\]
The middle and last inequalities have exact rational-power certificates
in Appendix~\ref{sec:constants}.

Let \(\beta_i\) be the coefficient in \eqref{eq:ordinary-off} for
an off-owner slot: \(1/2\) for an ordinary owner and \(1/3\) for a
discounted owner. H\"older combines the three slot charges into
the upper bound
\begin{equation}
L\left(z+R+\frac{3-k}{2}\right)^r-\frac R2,
\label{eq:ordinary-envelope}
\end{equation}
where
\begin{equation}
L=
\left(k+\frac{3-k-m}{2}
+\frac m2\cdot(2/3)^{12/5}\right)^{5/12},
\qquad \Lambda=\frac32-\frac m6.
\label{eq:ordinary-coefficient}
\end{equation}

Two elementary ways to compare this bound with the available budget will be used below.

\paragraph{Endpoint comparison.}
For a fixed shift \(b\ge1\) and a constant $C$, the difference \(L(z+b)^r-\Lambda(z+1)^r-C\) has its maximum on a compact interval at an endpoint. Its derivative is nonpositive if \(L\le \Lambda\).
Otherwise the derivative has at most one sign change, from negative to positive, so the maximum occurs at an endpoint. The derivative calculation is given in \Appendixref{sec:mechanical}.
For \(0\le b<1\), the difference is nonpositive if \(L\le \Lambda\) and $C$ is nonnegative; if \(L>\Lambda\), it increases. Thus the inequality either holds directly or reduces to checking the upper endpoint.

\paragraph{Marginal comparison when every owned slot has one unit.}
If \(R=k\), all the owned terms have the form \((1+x_i)^r\).
Let \(\Phi(z)\) be the largest possible value of
\[
\sum_{i\le k}(1+x_i)^r+
\sum_{i>k}\beta_i(1+2x_i)^r,
\qquad x_1+x_2+x_3=z.
\]
The bound needed is \(\Phi(z)-k/2\le \Lambda(1+z)^r\).

At a maximizing allocation with \(z>0\), every owned slot receives positive flow. Its initial marginal rate is \(r\), while the other slots start at rates at most \(r\); all rates decrease with flow.
A slot is active if $x_i>0$ at the maximizing allocation.
Write each active term as $b_i(x_i+d_i)^r$.
For an owned slot, $b_i=1$ and $d_i=1$; for an off-owner
slot, $b_i=2^r\beta_i$ and $d_i=1/2$.
Define
\[
\ell=\left(\sum_{i\text{ active}}b_i^{12/5}\right)^{5/12},
\qquad
d=\sum_{i\text{ active}}d_i.
\]
Thus \(\ell\le L\) and \(d\ge k\). Equal marginal rates give
\begin{equation}
\Phi'(z)=r\ell(z+d)^{-5/12}.
\label{eq:marginal}
\end{equation}
To see the formula, observe that equal derivatives make \(x_i+d_i\) proportional to \(b_i^{12/5}\), and their sum is \(z+d\). Strict concavity gives a unique maximizing allocation. At a change of the positive coordinates, the new coordinate has exactly the existing marginal rate, so the one-sided derivatives agree. There are at most three coordinates. Thus the derivative formula can be integrated across these changes as well as on each interval between them.

Consequently it is enough to check
\begin{equation}
\ell^{12/5}\le \Lambda^{12/5}\frac{z+d}{z+1}.
\label{eq:marginal-check}
\end{equation}
Then \(\Phi'(z)\le \Lambda r(z+1)^{-5/12}\).
Integrating from zero proves the desired bound, since
\(\Phi(0)-k/2=\Lambda\).

\paragraph{No owned slot: \(k=0\).}
Here \(\Lambda=\sum_i\beta_i\ge1\) and \(\beta_i\le1/2\).
Weighted concavity gives
\[
\sum_i\beta_i(1+2x_i)^r
\le \Lambda\left(1+\frac{2\sum_i\beta_i x_i}{\Lambda}\right)^r
\le \Lambda(1+z/\Lambda)^r\le \Lambda(1+z)^r.
\]

\paragraph{One owned slot: \(k=1\).}
For \(m=0,1,2\), respectively,
\[
\begin{array}{c|ccc}
m & 0 & 1 & 2 \\ \hline
\text{Upper bound on } L^{12/5}
  & 2 & \frac{17}{10} & \frac75 \\[4pt]
\Lambda
  & \frac32 & \frac43 & \frac76
\end{array}
\]
Here $(2/3)^{12/5}\le2/5$ gives the stated upper bounds.
In each column $L\le\Lambda$: for the first two, use
$2<(3/2)^2$ and $17/10<(4/3)^2$; for the third, Bernoulli's
inequality gives $(7/6)^{12/5}\ge1+(12/5)(1/6)=7/5$.
If \(R=0\), \eqref{eq:ordinary-envelope} directly gives the goal.
If \(R=1\), use the marginal comparison:
\(\ell\le L\le \Lambda\) and \(d\ge1\) prove
\eqref{eq:marginal-check} for every \(z\).

\paragraph{Two owned slots, ordinary third owner: \(k=2,m=0\).}
Here $L=(5/2)^{5/12}\le3/2=\Lambda$, since
$(5/2)^5\le(3/2)^{12}$.
For \(R=0\), the envelope immediately proves the goal.
For \(R=1,2\), its difference from \(\Lambda(1+z)^r\) decreases with \(z\),
so check only \(z=0\).
Weighted AM--GM gives, for $R=1,2$,
\[
(5/2)^{5/12}(R+1/2)^r-R/2
\le\frac5{12}\cdot\frac52+\frac7{12}(R+1/2)-\frac R2
=\frac43+\frac R{12}\le\frac32.
\]

\paragraph{Two owned slots, discounted third owner: \(k=2,m=1\).}
This case requires a separate endpoint estimate.
Here \(\Lambda=4/3\) and
\(L^{12/5}=2+(2/3)^{12/5}/2\).
Because the current center is ordinary, its normalized budget satisfies
\[
1+z\le(7/3)^{12/5}.
\]
For this paragraph only, call the upper endpoint
\[
T=(7/3)^{12/5}.
\]
Thus \(0\le z\le T-1\).

For $R=0,1$, the endpoint comparison applies. At $z=0$, put
$u=(2/3)^{12/5}\le2/5$, so $L^{12/5}=2+u/2\le11/5$.
Weighted AM--GM yields
\[
L(R+1/2)^r-R/2
\le\frac5{12}\cdot\frac{11}{5}+\frac7{12}(R+1/2)-\frac R2
=\frac{29}{24}+\frac R{12}\le\frac{31}{24}<\frac43.
\]
At the other endpoint, the rational bounds
\[
 L\le\frac{140}{101},\qquad 7<T\le\frac{153}{20},
 \qquad 6^{5/12}\ge2
\]
follow from the integer-power certificates in
Appendix~\ref{sec:constants}. For $0\le R\le1$, the derivative of
$f(R)=L(T-1/2+R)^r-R/2$ is negative: since $L<7/5$ and
$T-1/2>6$,
\[
 f'(R)\le\frac75\cdot\frac7{12}\cdot\frac12-\frac12<0.
\]
It therefore suffices to check $R=0$. The tangent bound at $1$ gives
\[
\frac{L(T-1/2)^r}{T^r}
\le\frac{140}{101}\left(1-\frac7{24T}\right)
\le\frac{140}{101}\left(1-\frac7{24}\cdot\frac{20}{153}\right)
=\frac{61810}{46359}<\frac43.
\]
This proves both upper-endpoint comparisons without numerical
approximations.

For \(R=2\), both owned slots contain a unit residual.
The coarse envelope loses too much at \(z=0\), so use the marginal
comparison instead. Now $z+1\le T<8$ and $\Lambda^{12/5}\ge28/15$;
the latter follows from $(4/3)^{12}\ge(28/15)^5$. There are only two active-slot possibilities:
\[
\begin{array}{c|cc}
\text{Third slot active?} & \ell^{12/5} & d \\ \hline
\text{No} & 2 & 2 \\[2pt]
\text{Yes} & 2+\frac12(2/3)^{12/5} & \frac52
\end{array}
\]
For the first row, the right side of \eqref{eq:marginal-check}
is at least
\[
\frac{28}{15}\left(1+\frac18\right)=\frac{21}{10}>2.
\]
For the second, it is at least
\[
\frac{28}{15}\left(1+\frac{3}{16}\right)
=\frac{133}{60}
> 2+\frac12(2/3)^{12/5}
.
\]
For the last strict inequality one may use
$(2/3)^{12/5}\le2/5$, giving $2+(2/3)^{12/5}/2\le11/5<133/60$.
Both rows satisfy the test.

\paragraph{Three owned slots: \(k=3\).}
Here $L=3^{5/12}$, $\Lambda=3/2$, and
\[
0\le z\le T-1,\qquad T=(8/3)^{12/5}.
\]
For each $R\in\{0,1,2,3\}$, the endpoint comparison reduces
the desired inequality to its values at $z=0$ and $z=T-1$.
At $z=0$, weighted AM--GM gives, for every $0\le R\le3$,
\[
3^{5/12}R^r-R/2
\le\frac5{12}\cdot3+\frac7{12}R-\frac R2
=\frac54+\frac R{12}\le\frac32.
\]
For the upper endpoint, use the exact bounds
\[
 L\le\frac{19}{12},\qquad T\le\frac{32}{3},
 \qquad T-1>6,\qquad 6^{5/12}\ge2.
\]
The first two are certified in Appendix~\ref{sec:constants}; the third
follows from $T>(8/3)^2=64/9>7$.
Consequently $f(R)=L(T-1+R)^r-R/2$ is decreasing for $R\ge0$, since
\[
 f'(R)\le\frac{19}{12}\cdot\frac7{12}\cdot\frac12-\frac12
 =-\frac{11}{288}<0.
\]
It remains to check $R=0$. Again using the tangent at $1$,
\[
 \frac{L(T-1)^r}{T^r}
 \le\frac{19}{12}\left(1-\frac7{12T}\right)
 \le\frac{19}{12}\left(1-\frac7{12}\cdot\frac3{32}\right)
 =\frac{2299}{1536}<\frac32.
\]
Thus all four upper endpoints satisfy the bound.

\subsection{Restoring scale and zero residuals}

Multiplying the local bound by \(t_a^r\) returns precisely the
account bound in Lemma~\ref{lem:account}.
This is the only rescaling used in the proof.

If \(t_a=0<c_a\), every other residual of an owned comparison vertex
is also zero, because its owner has maximum residual.
Apply the local inequality with an artificial residual
\(\epsilon>0\) and budget \(c_a+\epsilon\) at the current vertex,
keep all actual transfers and the other residuals fixed, and let
\(\epsilon\) decrease to zero.
The center's discount status is eventually constant; neighboring
owners' statuses are arbitrary fixed inputs to the local bound.
Every account entry converges to its required value.
If \(c_a=t_a=0\), all transfers and all entries in the account vanish.

This completes Lemma~\ref{lem:account}, and therefore the proof of
Theorem~\ref{thm:mixed}.

\section{Derivative and constant checks}
\label{sec:mechanical}

This appendix records the derivative arguments used in
\Cref{sec:mixed-proof} and \Appendixref{sec:local}.

\paragraph{Residual-only bound.}
For $0<x\le1$ and $0<r<1$, define
\[
h(x)=\frac{(1+x)^r-1}{x^r}.
\]
Differentiation gives
\[
h'(x)=r x^{-r-1}\bigl(1-(1+x)^{r-1}\bigr)\ge0.
\]
Hence $h(x)\le h(1)=2^r-1$, proving the residual inequality used
at the start of \Cref{sec:mixed-proof}. Moreover,
$\lim_{x\downarrow0}h(x)=0$.

\paragraph{Monotonicity of the increment.}
For $x,t>0$,
\[
\frac{\partial J_t(x)}{\partial t}
=r\bigl((t+2x)^{r-1}-(t+x)^{r-1}\bigr)\le0.
\]
Continuity extends this monotonicity to $t=0$ or $x=0$. In
particular,
\[
J_1(x)\le J_0(x)=(2^r-1)x^r.
\]

\paragraph{Endpoint comparison.}
For $b\ge1$, let
\[
D(z)=L(z+b)^r-\Lambda(z+1)^r-C,
\]
where $L,\Lambda>0$ and $C\ge0$. The sign of $D'(z)$ is the sign of
\[
\frac{L}{\Lambda}
\left(\frac{z+b}{z+1}\right)^{r-1}-1.
\]
Since $(z+b)/(z+1)$ is nonincreasing and $r-1<0$, this expression
is nondecreasing. Thus $D'$ can change sign only from negative to
positive, so the maximum of $D$ on a compact interval occurs at an
endpoint.

If $0\le b<1$ and $L\le\Lambda$, then $D(z)\le0$ directly. If
$0\le b<1$ and $L>\Lambda$, then $D'(z)>0$, so it suffices to
check the upper endpoint. These are the cases used in
\Appendixref{sec:ordinary}.

\subsection{Exact constant certificates}\label{sec:constants}
For completeness, the following table certifies the rational-power
bounds used above. In each row the right column is an equivalent
inequality between rational numbers raised to integer powers; clearing
denominators gives an integer comparison. Thus the verification uses
no approximation of irrational numbers.
\begin{center}
\renewcommand{\arraystretch}{1.22}
\begin{tabular}{@{}ll@{}}
\toprule
Bound used & Integer-power certificate\\
\midrule
$2^{7/12}\le3/2$ & $2^{19}\le3^{12}$\\
$(7/3)^{7/5}>5/2$ & $(7/3)^7>(5/2)^5$\\
$3^{5/12}\le8/5$ & $3^5\le(8/5)^{12}$\\
$(7/20)^{12/5}\le81/1000$ & $(7/20)^{12}\le(81/1000)^5$\\
$(1081/1000)^{5/12}\le31/30$
 & $(1081/1000)^5\le(31/30)^{12}$\\
$(11/20)^{12/5}\le6/25$ & $(11/20)^{12}\le(6/25)^5$\\
$(7/5)^{12/5}\ge56/25$ & $(7/5)^{12}\ge(56/25)^5$\\
$(7/16)^{12/5}\le7/50$ & $(7/16)^{12}\le(7/50)^5$\\
$(5/6)^{12/5}\ge16/25$ & $(5/6)^{12}\ge(16/25)^5$\\
$(2/3)^{12/5}\le2/5$ & $(2/3)^{12}\le(2/5)^5$\\
$(5/2)^{5/12}\le3/2$ & $(5/2)^5\le(3/2)^{12}$\\
$(2/3)^{12/5}\le189/500$ & $(2/3)^{12}\le(189/500)^5$\\
$(2189/1000)^{5/12}\le140/101$
 & $(2189/1000)^5\le(140/101)^{12}$\\
$(7/3)^{12/5}>7$ & $7^7>3^{12}$\\
$(7/3)^{12/5}\le153/20$ & $(7/3)^{12}\le(153/20)^5$\\
$6^{5/12}\ge2$ & $6^5\ge2^{12}$\\
$(4/3)^{12/5}\ge28/15$ & $(4/3)^{12}\ge(28/15)^5$\\
$3^{5/12}\le19/12$ & $3^5\le(19/12)^{12}$\\
$(8/3)^{12/5}\le32/3$ & $(8/3)^{12}\le(32/3)^5$\\
\bottomrule
\end{tabular}
\end{center}
In particular, the bound on the two-owned-slot coefficient follows as
\[
 L^{12/5}=2+\tfrac12(2/3)^{12/5}
 \le2+\tfrac12\cdot\tfrac{189}{500}
 =\tfrac{2189}{1000}\le(140/101)^{12/5}.
\]
The endpoint estimates above then use only concavity and these rational
bounds. They do not rely on a numerical optimization or a search over
real parameters.

\section{A locality lower bound for the two-phase scheme}
\label{sec:framework-lower-bound}

The following construction uses the familiar high-girth obstruction
for bounded-exchange local search~\cite{HS89,neuwohner2024}. We spell it
out to show that the gain-recovery phase does not remove this
obstruction. It gives a limitation under an allowed execution order,
not an approximation-hardness result or a lower bound for every
possible pivot rule.

\begin{proposition}\label{prop:two-stage-lower-bound}
For every fixed insertion bound $b\ge3$ and every power $p>0$, there
is a unit-weight $3$-set packing instance with an allowed first-phase
execution whose output is unchanged by the gain-recovery phase and
has weight exactly $2/3$ of optimum. The same statement holds with
the rounded first-phase potential.
\end{proposition}
\begin{proof}
Take a finite simple $3$-regular graph $H=(V,E)$ of girth greater than
$b$. Introduce one ground element $(v,e)$ for each incidence of a
vertex and an edge. For $v\in V$ and $e=uv\in E$, define
\[
 A_v=\{(v,e):e\ni v\},\qquad
 O_e=\{(u,e),(v,e)\}.
\]
Let $A=\{A_v:v\in V\}$ and $O=\{O_e:e\in E\}$, and give every set
weight one. Both $A$ and $O$ are packings, and their sets have sizes
three and two, respectively. Their conflict graph is the incidence
graph of $H$.

Every nonempty $X\subseteq O$ with $|X|\le b$ corresponds to an
edge-induced forest in $H$. If it has $c_X\ge1$ nonempty components,
then
\[
 |\N_A(X)|=|X|+c_X>|X|.
\]
All unit weights have the same positive $p$-power contribution.
Therefore no such insertion improves the first-phase potential.
Since $b\ge3$, every candidate of the recovery phase also has strictly
negative original-weight gain. An allowed execution first inserts
all members of $A$ as singletons and then stops at $A$; recovery
leaves it unchanged. For the rounded potential all sets again have
the same positive integer contribution, so the argument is identical.

To identify the optimum, consider a packing that contains the current
sets indexed by $T\subseteq V$. The number of edge sets excluded by
these choices is
\[
 3|T|-|E(H[T])|\ge\tfrac32|T|.
\]
Its total cardinality is consequently at most
$|T|+|E|-\tfrac32|T|\le|E|$. Thus $O$ is optimal, and regularity gives
\[
 \frac{w(\OPT)}{w(A)}=\frac{|E|}{|V|}=\frac32.
\]
\end{proof}

The proposition applies for each fixed accuracy, because the insertion
bound of Algorithm~\ref{alg:two-stage} is then a constant. It rules out
a factor strictly below $3/2$ for a guarantee uniform over all allowed
first-phase choices in this two-stage framework. It does not rule out
an exact $3/2$ guarantee, an algorithm with larger or different
neighborhoods, or a stronger guarantee obtained by imposing additional
requirements on the pivot rule.

\section{Counterexamples for other power exponents}
\label{sec:exponent-examples}

We show that some power exponents fail even after the recovery
phase. Replace $12/7$ by a fixed $p>0$, keeping the insertion
bound $b=4^{\lceil1/\eps\rceil}-1$ and the recovery phase
unchanged. We first analyze the unrounded potential
$\Phi_p(Y)=\sum_{S\in Y}w(S)^p$ and address rounding at the end.

\subsection{Small powers: \texorpdfstring{$0<p<\log_{3/2}2$}{0 < p < log(2)/log(3/2)}}

Choose a rational number
\[
\frac32<\beta<\min\{2,2^{1/p}\}.
\]
Take a finite simple $3$-regular bipartite graph with sides
$A,O$ and girth greater than $2\max\{b,3\}$.
Such graphs exist with arbitrarily large girth; this is the standard
high-girth setting for local-search lower bounds~\cite{HS89,neuwohner2024}.
Represent each vertex by the set of its three incident edges,
so that both sides are packings and adjacency corresponds
to intersection. Give sets in $A$ weight $1$ and sets in
$O$ weight $\beta$. Regularity gives $|A|=|O|$.

For every nonempty $X\subseteq O$ with
$k=|X|\le\max\{b,3\}$, the graph on $X\cup\N_A(X)$ is a forest
with $3k$ edges. Hence
\[
|\N_A(X)|\ge2k+1.
\]
Since $\beta^p<2$ and $\beta<2$,
\[
\Phi_p(X)<2k<\Phi_p(\N_A(X)),
\qquad
w(X)<2k<w(\N_A(X)).
\]
Thus $A$ admits neither a permitted potential improvement
nor a positive-gain recovery candidate.

An allowed execution builds $A$ by giving its singleton
insertions priority and then returns $B=A$. However,
\[
\frac{w(\OPT)}{w(B)}
\ge\frac{w(O)}{w(A)}
=\beta>\frac32.
\]
The gap depends only on $p$, not on the insertion bound.
This rules out a guarantee uniform over the permitted
first-phase choices.

\subsection{Large powers: \texorpdfstring{$p\ge1.83$}{p >= 1.83}}

Start with seven disjoint stars.
The distinguished center has weight $1$ and three leaves
of weight $0.5479$ each. Each of the other six centers has
weight $1.01$ and three leaves of weights
\[
0.64236,\qquad 0.505,\qquad 0.505.
\]
All terminating decimals denote exact rational numbers.

Add three mutually disjoint sets $d_1,d_2,d_3$, each of
weight $0.548$. Each $d_j$ intersects the distinguished
center, its $j$th leaf, and the heavy leaves of outer centers
$2j-1$ and $2j$, with no other intersections.
These intersections can be realized by sets of size at most
three: use one ground element for each triangle formed by
the distinguished center, its $j$th leaf, and $d_j$, and
one separate element for every remaining edge of the
specified intersection graph. Each vertex uses at most
three elements.

Let $A$ be the seven centers and $D=\{d_1,d_2,d_3\}$.
Every noncenter set meets exactly one center.
Relative to its center's power, the alternatives at an
outer center contribute at most
\[
0.636^p+2(1/2)^p,
\]
while those at the distinguished center contribute at most
$3(0.548)^p$.
Direct calculation gives
\[
0.636^{1.83}+2(1/2)^{1.83}<1,
\qquad
3(0.548)^{1.83}<1.
\]
For an exact rational verification at the endpoint, use
\[
 (159/250)^{183}<(43685/100000)^{100},\qquad
 (1/2)^{183}<(28127/100000)^{100},
\]
\[
 (137/250)^{183}<(33264/100000)^{100}.
\]
These are integer comparisons after clearing denominators. Taking
$100$th roots gives an outer sum below $99939/100000<1$
and a distinguished-center sum below $99792/100000<1$.
Both expressions decrease with $p$.
Thus every center contributes strictly more power than
any packing of its alternatives. If $b\ge7$, inserting
all missing centers improves every packing other than $A$.
Consequently every first-phase execution ends at $A$.

In the recovery phase, the three outer leaves at any one
center have gain
\[
0.64236+2(0.505)-1.01=0.64236,
\]
whereas $D$ has gain
\[
3(0.548)-1=0.644.
\]
No candidate involving several centers has positive gain:
its inserted weight is at most $3(0.64236)<2$, while its
deletion cost is at least $2$.
Among candidates involving only the distinguished center,
$D$ uniquely maximizes the gain.
The greedy procedure therefore accepts $D$ first.

This blocks all three distinguished leaves and all six
heavy outer leaves. The two remaining leaves at each
outer center have combined weight equal to that center,
so no further compatible positive-gain candidate is
accepted. Hence
\[
w(B)=6(1.01)+3(0.548)=7.704.
\]
All twenty-one original leaves form a packing $O$ with
\[
w(O)=3(0.5479)+6\bigl(0.64236+2(0.505)\bigr)
=11.55786.
\]
Therefore
\[
\frac{w(\OPT)}{w(B)}
\ge\frac{w(O)}{w(B)}
=\frac32+\frac{31}{128400}.
\]

\paragraph{Rounding and consequences.}
Both constructions remain valid for
$Q_p(S)=\lfloor(w(S)/\tau)^p\rfloor$.
For fixed $p$ and $\eps$, the strict potential comparisons
above have positive margins.
Taking sufficiently many disjoint copies decreases
$\tau=\eps M/(30n)$ without changing the weights or $b$.
Each required comparison involves a bounded number of
sets, so its margin, scaled by $\tau^{-p}$, eventually
exceeds the additive rounding error.
The first-phase arguments therefore remain valid.
The recovery phase uses the original weights; its behavior
is unchanged, and candidates involving different copies
have negative gain by the same bounds used above.

For sufficiently small $\eps$, we have $b\ge7$.
The first construction violates the target guarantee when
$\eps<\beta-3/2$, and the second does so when
$\eps<31/128400$.
Thus a guarantee valid for every fixed accuracy and every
permitted first-phase order requires
\[
\log_{3/2}2\le p<1.83.
\]
These examples do not establish sufficiency throughout
this interval.

\end{document}